\documentclass{article}

\usepackage{microtype}
\usepackage{graphicx}
\usepackage{subcaption}
\usepackage{booktabs}
\usepackage{hyperref}
\usepackage{multirow}

\usepackage[preprint]{icml2026}

\usepackage{amsmath}
\usepackage{amssymb}
\usepackage{mathtools}
\usepackage{amsthm}
\usepackage[capitalize,noabbrev]{cleveref}
\usepackage[disable,textsize=tiny]{todonotes}

\usepackage{cuted}
\usepackage{capt-of}

\usepackage{stfloats}
\newcommand{\molexar}{Molexar}
\newcommand{\fselfies}{Fragment-SELFIES}
\usepackage[table]{xcolor}
\definecolor{macaronblue}{HTML}{A7C7E7}    
\definecolor{macaronred}{HTML}{FFAAA5}     
\definecolor{macaronpurple}{HTML}{CDB4DB}  
\newcommand{\colhead}{Method & Valid. & Uniq. & Div. & QED & SA & Lip. & Vina & HA}

\icmltitlerunning{\molexar{}: A Unified Multimodal Molecular Foundation Model for Drug Design}

\begin{document}

\twocolumn[
  \icmltitle{\molexar{}: A Unified Multimodal Molecular Foundation Model for Drug Design}

  \icmlsetsymbol{equal}{*}

  \begin{icmlauthorlist}
    \icmlauthor{Haoyu Lin}{equal,cqb}
    \icmlauthor{Yiyan Liao}{equal,sls}
    \icmlauthor{Jinmei Pan}{pathology}
    \icmlauthor{Xinliao Ling}{cqb}
    \icmlauthor{Luhua Lai}{cqb,bnlms,ptcls,pucd}
    \icmlauthor{Jianfeng Pei}{cqb}
  \end{icmlauthorlist}

  \icmlaffiliation{cqb}{Center for Quantitative Biology, Academy for Advanced Interdisciplinary Studies, Peking University, Beijing 100871, China}
  \icmlaffiliation{sls}{School of Life Sciences, Peking University, Beijing 100871, China}
  \icmlaffiliation{pathology}{Department of Pathology, The First Affiliated Hospital of Shantou University Medical College, Shantou, Guangdong 515041, China}
  \icmlaffiliation{bnlms}{BNLMS, College of Chemistry and Molecular Engineering, Peking University, Beijing 100871, China}
  \icmlaffiliation{ptcls}{Peking-Tsinghua Center for Life Sciences, Academy for Advanced Interdisciplinary Studies, Peking University, Beijing 100871, China}
  \icmlaffiliation{pucd}{Peking University Chengdu Academy for Advanced Interdisciplinary Biotechnologies, Chengdu, Sichuan 610213, China}

  \icmlcorrespondingauthor{Luhua Lai}{lhlai@pku.edu.cn}
  \icmlcorrespondingauthor{Jianfeng Pei}{jfpei@pku.edu.cn}

  \icmlkeywords{molecular generation, foundation models, multimodal conditioning, Fragment-SELFIES, structure-based drug design}

  \vskip 0.3in
]

\printAffiliationsAndNotice{\icmlEqualContribution}

\begin{abstract}
Molecular generation is a central challenge in drug discovery, requiring models that explore vast chemical space while satisfying diverse design constraints. We present \molexar{}, a unified multimodal molecular foundation model built on \fselfies{}, a robust, fragment-aware molecular language with validity-preserving decoding and explicit fragment structure. A pretrained autoregressive decoder learns the \fselfies{} syntax and molecular distribution; supervised fine-tuning (SFT) then trains the same decoder on condition-molecule pairs spanning scalar molecular properties, pharmacophore fingerprints, protein sequences, and binding pockets, injecting each condition by in-place replacement of value-token embeddings so that all generation modes share one autoregressive path. \molexar{} achieves strong efficiency at a small parameter count while matching or exceeding larger models. The pretrained model reaches 100\% validity and high drug-likeness in unconditional and fragment-constrained generation; the SFT model follows single- and multi-property instructions and remains competitive on target-conditioned generation on the CrossDocked2020 test set. On MolGenBench, \molexar{} further generates molecules with favorable safety and potency. These results establish \molexar{} as a practical unified foundation for computational chemistry and drug-design workflows.

\end{abstract}

\section{Introduction}

Generative modeling has become a central component of modern drug discovery because it promises to search chemical spaces that are far larger than experimentally accessible libraries. Early molecular language models and variational or adversarial generators demonstrated that strings and graphs can be optimized toward desired molecular properties, including GPT-style SMILES generation with property and scaffold conditioning \cite{Bagal2022MolGPT}. Practical molecular design, however, rarely involves a single objective. A useful generator must simultaneously produce valid molecules, preserve or modify fragments, control scalar properties, satisfy pharmacophore hypotheses, and exploit target information from protein sequences or three-dimensional binding pockets. These requirements have motivated a broad range of specialized models, including chemical language models for property-directed design, structure-based generators for protein pockets, pharmacophore-conditioned models, and fragment-aware representations.

Molecular representation is a recurring bottleneck. SMILES is compact and widely used, but not every token sequence maps to a valid molecule \cite{Weininger1988SMILES}. SELFIES addresses this by defining a robust string representation in which arbitrary SELFIES strings decode to valid molecules \cite{Krenn2020SELFIES,Krenn2022SELFIESFuture}. More recent fragment-aware representations, including Group SELFIES, SAFE, and t-SMILES, explicitly encode chemically meaningful substructures to improve generation, optimization, and fragment-constrained design \cite{Cheng2023GroupSELFIES,Noutahi2024SAFE,Wu2024TSMILES}. Fragment-level representations are particularly attractive for medicinal chemistry because scaffold decoration, linker design, and hit-to-lead optimization are often expressed as operations on fragments rather than isolated atoms.

At the same time, model architectures have diverged across design settings. Molecular foundation models such as ChemFM, GenMol, and SmileyLlama explore the use of large language or diffusion models for broad molecular generation tasks \cite{Cai2025ChemFM,Lee2025GenMol,Cavanagh2026SmileyLlama}. Protein- or target-aware models such as TamGen introduce target conditioning into chemical language models \cite{Wu2024TamGen}. In structure-based drug design, receptor-conditioned generators have evolved from direct 3D ligand construction methods such as DeepLigBuilder to grid-based models, autoregressive equivariant models, flow-based pocket generators, and diffusion models conditioned on protein pockets \cite{Li2021SBDD,Ragoza2022ReceptorCVAE,Luo2022SBDD,Liu2022GraphBP,Peng2022Pocket2Mol,Zhang2023ResGen,Jiang2024PocketFlow,Guan2023TargetDiff,Schneuing2024DiffSBDD,Huang2024PMDM,Guan2024DecompDiff,Huang2024IRDiff,Lin2025DiffBP,Qu2024MolCRAFT}. PocketXMol recently showed that an atom-level prompt and universal denoising framework can unify many pocket-interaction tasks, including docking, structure-based design, fragment operations, and peptide design \cite{Peng2026PocketXMol}. Other methods focus on pharmacophores, interactions, shapes, or electrostatics as design priors \cite{Imrie2021DEVELOP,Skalic2019LigDream,Zhu2023PGMG,Xie2025TransPharmer,Peng2025PhoreGen,Zhung2024DeepICL,Adams2025ShEPhERD,Wang2024PharmacoBridge}. While these systems have advanced individual tasks, their heterogeneity makes it difficult to share representations, transfer pretraining, or compare conditioning strategies across modalities.

Evaluation has likewise become more demanding. General molecular generation benchmarks such as GuacaMol and MOSES emphasize distribution learning, validity, novelty, and goal-directed optimization \cite{Brown2019GuacaMol,Polykovskiy2020MOSES}. Structure-based evaluation depends on curated protein-ligand resources such as PDBbind, CASF, and CrossDocked2020 \cite{Wang2004PDBbind,Liu2015PDBbind,Li2018CASF2013,Su2019CASF2016,Francoeur2020CrossDocked}, while virtual-screening benchmarks such as DUD-E and DEKOIS 2.0 test enrichment against challenging decoys \cite{Mysinger2012DUDE,Bauer2013DEKOIS}. Recent PoseBusters and PoseCheck analyses show that docking or RMSD scores alone can hide chemically invalid or strained generated poses \cite{Buttenschoen2024PoseBusters,Harris2023PoseCheck}. Newer resources including Durian, POKMOL-3D, PLINDER, SAIR, and MolGenBench further emphasize leakage-aware splits, realistic protein-ligand systems, active recovery, and hit-to-lead applicability \cite{Nie2025Durian,Liu2024POKMOL3D,Durairaj2024PLINDER,Lemos2025SAIR,Cao2025MolGenBench}. These trends motivate a model interface that can support multiple evaluation regimes without rebuilding the generator for each modality.

We present \molexar{}, a unified multimodal molecular foundation model designed to reuse one autoregressive molecular decoder across these settings. \molexar{} is based on the Gemma2 architecture \cite{GemmaTeam2024Gemma2}; however, its molecular language, condition interface, and training objective are specialized for molecular design. The model operates on \fselfies{}, a BRICS-fragment molecular representation developed for this work that uses an explicit fragment-tree token grammar rather than corpus-specific fragment identifiers. \fselfies{} encodes the BRICS fragment tree with a small set of structural tokens, represents fragment interiors using SELFIES-compatible tokens, and exposes dummy attachment sites that support fragment-constrained prompts. This design retains chemically meaningful fragment structure while avoiding opaque fragment IDs. The complete codec and model workflow are summarized in \cref{fig:fragment-selfies-codec,fig:molexar-architecture}.

The main design goal of \molexar{} is to make conditional molecular generation a prompt-embedding problem rather than an architecture-specific problem. All conditions are inserted into a fixed template as token-like values. During the forward pass, embeddings at selected value-token positions are replaced by projections of condition values. Discrete properties are encoded by one-hot embeddings, continuous properties by radial basis expansions, vector conditions by learned projections, and pocket graphs by a GVP encoder \cite{Jing2021GVP}. This strategy allows unconditional pretraining, conditional supervised fine-tuning, and fragment continuation to share the same causal language modeling path while preserving key-value-cache-compatible generation.

This paper makes the following contributions. First, we introduce \fselfies{}, a fragment-level molecular representation with an explicit fragment-tree token grammar that avoids corpus-specific fragment-ID tokens while supporting foundation-model pretraining and fragment-constrained generation. Second, we describe \molexar{}, a Gemma2-derived molecular language model that unifies unconditional and multimodal conditional generation through value-token embedding replacement. Third, we evaluate the same model across unconditional generation, inference efficiency, fragment-constrained design, property control, target-conditioned generation on CrossDocked2020 test set, and real-world applicability on MolGenBench, comparing against molecular language models, pharmacophore-guided generators, and structure-based drug design systems.

\section{Related Work}

\subsection{Molecular String and Fragment Representations}

SMILES remains the dominant textual molecular representation due to its compactness and software support \cite{Weininger1988SMILES}. Its fragility under arbitrary sequence perturbations, however, can make language-model decoding inefficient when validity is important. SELFIES was introduced to address this limitation by making molecular strings robust under arbitrary token sequences \cite{Krenn2020SELFIES,Krenn2022SELFIESFuture}. Group SELFIES extends the same philosophy with group-level tokens, allowing larger chemical units to be represented explicitly \cite{Cheng2023GroupSELFIES}. SAFE uses fragment strings connected by attachment labels and demonstrates strong performance in fragment-constrained generation and molecular design \cite{Noutahi2024SAFE}. t-SMILES further explores tree-like fragment decomposition for ligand design \cite{Wu2024TSMILES}. \fselfies{} follows this fragment-aware direction by explicitly serializing a BRICS fragment tree while encoding each fragment body with SELFIES-compatible tokens and non-atomic dummy attachment placeholders. This preserves chemically meaningful fragment structure for constrained prompts and avoids the fixed predefined group-token set required by Group SELFIES without representing fragments as opaque IDs.

\subsection{Molecular Foundation and Language Models}

MolGPT showed that a decoder-only Transformer trained by next-token prediction over SMILES can generate valid molecules and can be conditioned on scalar properties and Bemis-Murcko scaffolds \cite{Bagal2022MolGPT}. More recent work has increasingly treated molecules as a domain for larger foundation models. ChemFM studies scaling-law-guided chemical pretraining over molecular strings and supports property-related downstream tasks \cite{Cai2025ChemFM}. GenMol formulates a drug-discovery generalist through discrete diffusion over molecular sequences \cite{Lee2025GenMol}. SmileyLlama adapts large language models for directed chemical space exploration with natural-language-like property prompts \cite{Cavanagh2026SmileyLlama}. MolFM aligns molecular graphs, biomedical text, and knowledge-graph context for multimodal molecular representation learning \cite{Luo2023MolFM}. TamGen combines a protein encoder and chemical decoder to generate molecules conditioned on target information \cite{Wu2024TamGen}. These models show that molecular generation and representation learning can benefit from general sequence-modeling and multimodal pretraining machinery, but they often specialize their conditioning interface to a particular input form. \molexar{} instead uses a uniform value-token interface that can bind scalar properties, fingerprints, protein embeddings, and pocket embeddings to the same molecular decoder.

\subsection{Structure-Based and Feature-Guided Generation}

Structure-based drug design requires models to generate molecules that are chemically valid and compatible with a target binding site. Early deep generative approaches included DeepLigBuilder, which combines an end-to-end 3D ligand generator with Monte Carlo tree search to optimize molecules directly inside binding pockets, and grid- or density-based receptor-conditioned models \cite{Li2021SBDD,Ragoza2022ReceptorCVAE}. Subsequent models used equivariant or geometric networks for autoregressive atom placement, as in SBDD, GraphBP, Pocket2Mol, ResGen, and PocketFlow \cite{Luo2022SBDD,Liu2022GraphBP,Peng2022Pocket2Mol,Zhang2023ResGen,Jiang2024PocketFlow}. Diffusion-based models such as TargetDiff, DiffSBDD, PMDM, DecompDiff, IRDiff, and DiffBP generate full molecular structures conditioned on pocket geometry, decomposed priors, or reference-ligand interactions and have become important baselines for 3D SBDD \cite{Guan2023TargetDiff,Schneuing2024DiffSBDD,Huang2024PMDM,Guan2024DecompDiff,Huang2024IRDiff,Lin2025DiffBP}. PocketXMol broadens this geometric line by representing tasks with atom-level prompts and training a universal denoiser across small-molecule and peptide tasks, yielding a unified 3D model for pocket-based generation and docking \cite{Peng2026PocketXMol}. MolCRAFT emphasizes conformational quality and highlights that high docking scores can be misleading when generated poses are strained or unrealistic \cite{Qu2024MolCRAFT}. In parallel, DEVELOP, LigDream, PGMG, TransPharmer, PhoreGen, DeepICL, ShEPhERD, and PharmacoBridge demonstrate the value of pharmacophore, interaction, shape, electrostatic, and linker constraints \cite{Imrie2021DEVELOP,Skalic2019LigDream,Zhu2023PGMG,Xie2025TransPharmer,Peng2025PhoreGen,Zhung2024DeepICL,Adams2025ShEPhERD,Wang2024PharmacoBridge}. \molexar{} is complementary to these methods: it is a molecular language model, but its condition interface is designed to absorb the same categories of target and feature information.

\section{Methods}

\subsection{Fragment-SELFIES Molecular Language}

\begin{figure*}[!t]
  \centering
  \includegraphics[width=\textwidth]{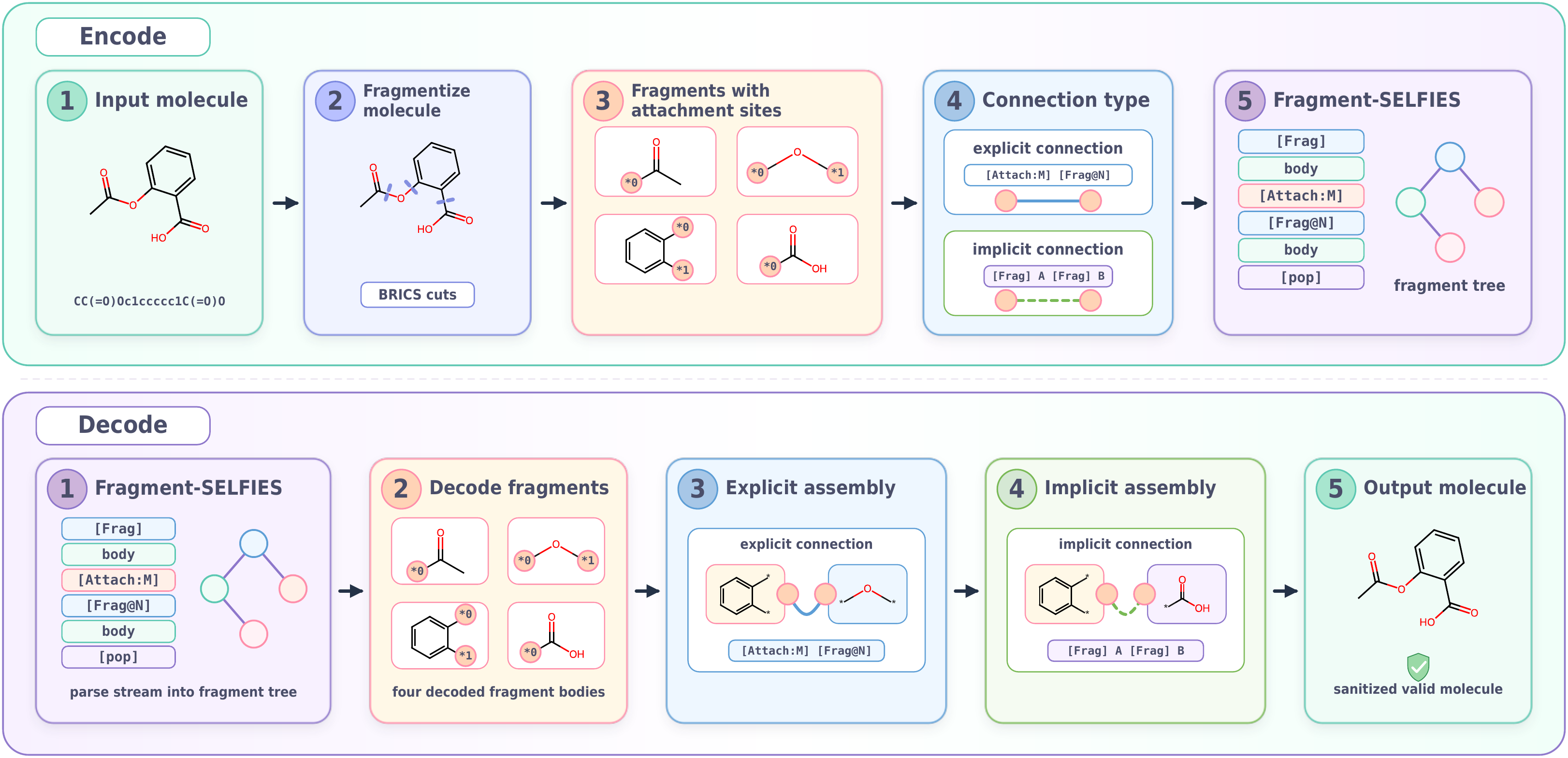}
  \caption{\textbf{\fselfies{} encoding and decoding workflow.} A molecule or SMILES string is BRICS-fragmented into chemically meaningful fragments with dummy attachment sites. The encoder serializes the resulting fragment tree with root \texttt{[Frag]} tokens, parent-side \texttt{[Attach:M]} selectors, child-side \texttt{[Frag@N]} tokens, and \texttt{[pop]} returns while keeping each fragment body in SELFIES-compatible tokens. The connection-type panel distinguishes explicit connections, where \texttt{[Attach:M]} and \texttt{[Frag@N]} define the parent and child attachment sites, from implicit connections, schematically \texttt{[Frag] A [Frag] B}, where adjacent root fragments do not specify the bonding relationship. During decoding, explicit edges are reconnected directly, whereas implicit anchors are assembled through compatible unused dummy attachment sites, making the implicit form useful as a linker-design start string. The fallback \texttt{[SELFIES]}...\texttt{[ENDSELFIES]} block preserves molecules that are not represented by the BRICS-fragment form.}
  \label{fig:fragment-selfies-codec}
  \vskip -0.1in
\end{figure*}

\fselfies{} represents a molecule as a BRICS fragment tree (\cref{fig:fragment-selfies-codec}). A root fragment begins with \texttt{[Frag]}. An explicit connection is encoded by \texttt{[Attach:M]} followed by \texttt{[Frag@N]}: \texttt{[Attach:M]} selects the parent-side attachment site, and \texttt{[Frag@N]} starts the child fragment while selecting its child-side attachment site. This explicitly specifies the bonding relationship between the two fragments; \texttt{[pop]} then returns traversal to the parent. The body of each fragment is encoded by SELFIES-compatible tokens such as atom, branch, and ring tokens. Attachment sites are represented by non-atomic \texttt{[Dummy]} tokens, which are removed or reconnected during decoding. A \texttt{[SELFIES]}...\texttt{[ENDSELFIES]} fallback can encode molecules for which the BRICS-fragment form is not used.

This representation is designed for molecular language modeling. First, it does not rely on a corpus-specific fragment-ID token set at encode/decode time: it combines a small fixed set of structural tokens with SELFIES-compatible fragment-body tokens. Second, BRICS fragments expose chemically meaningful units, making fragment-level prompts natural. Third, adjacent root fragments can form an implicit connection, written schematically as \texttt{[Frag] A [Frag] B}. Unlike an explicit connection of the form \texttt{[Attach:M] [Frag@N]}, this form does not prescribe which attachment sites should bond. It is therefore designed for linker-design start strings: the user can provide two endpoint fragments as the initial string, and generation can continue by proposing the missing linker fragments while decoding assembles compatible unused dummy attachment sites.

As a concrete example matching \cref{fig:fragment-selfies-codec}, aspirin, \texttt{CC(=O)Oc1ccccc1C(=O)O}, is decomposed into four BRICS fragments: an acetyl fragment, an ester oxygen, an aryl ring, and a carboxyl fragment. One equivalent \fselfies{} serialization is shown below, with line breaks added only for readability:
\begin{center}
\begin{minipage}{\linewidth}
\scriptsize\ttfamily
\noindent [Frag][C][CH0][=Branch1][C][=O][Dummy][Attach:0]\par
\noindent [Frag@0][Dummy][OH0][Dummy][Attach:1]\par
\noindent [Frag@0][Dummy][CH0][=C][C][=C][C]\par
\noindent [=CH0][Ring1][=Branch1][Dummy][pop][pop]\par
\noindent [Frag][O][=CH0][Branch1][C][O][Dummy]
\end{minipage}
\end{center}
The first three fragments are connected explicitly: acetyl attachment 0 connects to oxygen attachment 0, and oxygen attachment 1 connects to aryl attachment 0. The final carboxyl fragment starts with a new \texttt{[Frag]} root rather than an \texttt{[Attach:M] [Frag@N]} edge, so the aryl-carboxyl relationship is implicit. During decoding, the parser first reconstructs the explicit acetyl-oxygen-aryl tree, then assembles the adjacent carboxyl root through the remaining compatible dummy attachment sites. This string was checked with the \fselfies{} implementation and strict decoding recovers the canonical aspirin SMILES \texttt{CC(=O)Oc1ccccc1C(=O)O}.

\subsection{Molexar Sequence Template}

\begin{figure*}[!t]
  \centering
  \includegraphics[width=\textwidth]{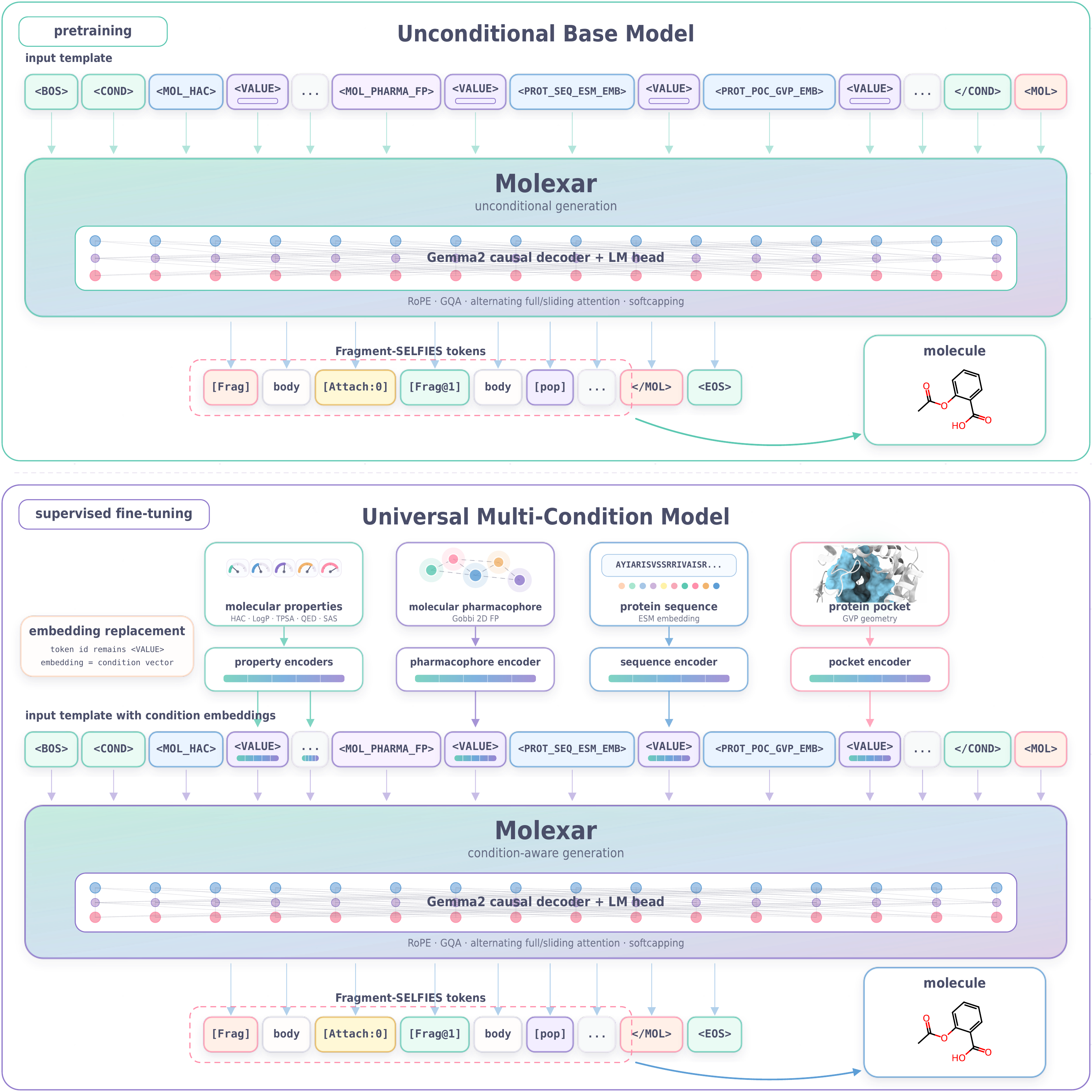}
  \caption{\textbf{\molexar{} architecture and training flow.} Stage 1 performs unconditional causal-language-model pretraining on \fselfies{} strings with the unified prompt template and a Gemma2-style molecular decoder. Stage 2 performs universal multi-condition supervised fine-tuning: scalar property encoders, pharmacophore-fingerprint projections, protein-sequence embeddings, and pocket-geometry encoders produce vectors that replace designated \texttt{<VALUE>} token embeddings in the condition block. The autoregressive decoder, molecular output block, attention masks, and key-value-cache-compatible generation path are shared across unconditional generation, property- and pharmacophore-guided generation, target-aware generation, pocket-conditioning generation, and fragment continuation.}
  \label{fig:molexar-architecture}
  \vskip -0.1in
\end{figure*}

\molexar{} uses a shared sequence template for pretraining, supervised fine-tuning, and inference:
\begin{center}
\small
\texttt{\textless BOS\textgreater\textless COND\textgreater\ conditions \textless /COND\textgreater\textless SEP\textgreater}\\
\texttt{\textless MOL\textgreater\ molecule \textless /MOL\textgreater\textless EOS\textgreater}.
\end{center}
The condition block contains condition keys and placeholder value tokens. The molecular block contains a \fselfies{} string. In unconditional pretraining, no condition value is active and the model learns the molecular grammar and distribution. In conditional fine-tuning and inference, selected placeholder value-token embeddings are replaced with encoded condition vectors, while all surrounding tokens and the autoregressive objective remain unchanged.

In the implementation, the condition block is an ordered sequence of key-token/value-token pairs. With zero-indexed token positions in the full template, the value placeholder for the $i$th condition slot is
\begin{equation}
  p_i = 3 + 2i, \qquad i=0,\ldots,21,
\end{equation}
where the first twelve slots correspond to molecular properties, pharmacophore fingerprint, protein-sequence embedding, and pocket-geometry embedding, and the remaining ten slots are reserved unused slots. Thus changing the active condition set changes only which fixed positions are replaced, not the token sequence length.

\subsection{Backbone Architecture}

The decoder is a Gemma2-style causal transformer with rotary positional embeddings, grouped-query attention, sliding-window/full-attention layer patterns, and logit softcapping inherited from the Gemma2 design \cite{GemmaTeam2024Gemma2}. As shown in \cref{fig:molexar-architecture}, we train it as a molecular causal language model over \fselfies{} tokens and reuse the same decoder for unconditional and conditional generation. The experimental model has 16 transformer layers, hidden size 256, intermediate size 640, four attention heads, one key-value head, maximum sequence length 256, sliding-window size 128, and a 127-token vocabulary. The tokenizer is a word-level tokenizer whose regular expression treats molecular and control tokens as atomic units. This keeps chemical tokens intact and the molecular token set interpretable.

Let $h_t^{(\ell)}$ denote the hidden state at position $t$ in layer $\ell$. For query head $a$ and its grouped key-value head $g(a)$, attention uses the allowed causal set $\mathcal{A}_\ell(t)$, which is either all positions up to $t$ in full-attention layers or the most recent 128 positions up to $t$ in sliding-window layers. With $\tau_{\mathrm{attn}}=50$,
\begin{align}
  \lambda_{ts}^{(\ell,a)} & =
  \tau_{\mathrm{attn}}\tanh\!\left(
  \frac{(q_t^{(\ell,a)})^\top k_s^{(\ell,g(a))}}
       {\tau_{\mathrm{attn}}\sqrt{d_h}}
  \right), \\
  \alpha_{ts}^{(\ell,a)} & =
  \operatorname{softmax}_{s\in\mathcal{A}_\ell(t)}
  \left(\lambda_{ts}^{(\ell,a)}\right).
\end{align}
The attention output is $\sum_{s\in\mathcal{A}_\ell(t)}\alpha_{ts}^{(\ell,a)}v_s^{(\ell,g(a))}$, followed by the Gemma2 feed-forward, normalization, and residual updates.

\subsection{Unified Condition Injection}

Let $x_{0:T-1}$ be the tokenized template and let $E(x_t)$ be the learned input embedding at position $t$. For an active condition $c_i$ with a designated value-token position $p_i$, \molexar{} first maps the raw condition to a feature vector
\begin{equation}
  r_i(c_i)=
  \begin{cases}
    \operatorname{onehot}_{\Omega_i}(c_i), & \text{discrete scalar},\\
    [\psi_{ij}(c_i)]_{j=1}^{K_i}, & \text{continuous scalar},\\
    c_i, & \text{direct vector condition},
  \end{cases}
\end{equation}
where $\Omega_i$ is the discrete allowable set, $\psi_{ij}(c_i)=\exp(-(c_i-\mu_{ij})^2/(2\sigma_i^2))$, and the RBF centers $\mu_{ij}$ are uniformly spaced over the configured property range. A two-layer projector then produces a hidden-size condition embedding
\begin{equation}
  z_i = P_i(r_i(c_i))
  = W_{i,2}\,\phi(W_{i,1}r_i(c_i)+b_{i,1}) + b_{i,2}
  \in \mathbb{R}^{d},
\end{equation}
with $d=256$ and the GELU-tanh activation $\phi$. The input embedding sequence is modified by
\begin{equation}
  \tilde{e}_t =
  \begin{cases}
    z_i, & t = p_i \text{ for an active condition } i, \\
    E(x_t), & \text{otherwise}.
  \end{cases}
\end{equation}
The transformer receives $\tilde{e}_{0:T-1}$ instead of the original embeddings. Missing conditions are assigned out-of-range positions by the collator and therefore do not replace any embedding. This operation does not add tokens, does not alter attention masks, and does not require cross-attention. Therefore, generation remains compatible with the standard autoregressive key-value cache.

\molexar{} currently supports twelve built-in condition types. Discrete scalar conditions include heavy atom count (HAC), hydrogen-bond donor count (HBDC), hydrogen-bond acceptor count (HBAC), and rotatable bond count (RotBC). Continuous scalar conditions include molecular weight (WT, Da), LogP, topological polar surface area (TPSA), QED, and synthetic accessibility score (SAS), encoded through radial basis functions before projection. Vector conditions include a 2D pharmacophore fingerprint, a protein sequence embedding, and a pocket-geometry embedding. The protein-sequence condition uses mean-pooled ESMC-600M final embeddings with dimension 1152 \cite{Candido2026ESMProtein}. Additional unused condition slots are reserved for user-defined conditions.

\subsection{Pocket Geometry Encoder}

For pocket-conditioned generation, a protein pocket is converted to an atom graph. We parse no-hydrogen pocket PDB files, keep atoms within a 25\,\AA{} radius of the pocket center, cap each pocket at 425 atoms, and build a directed 8-nearest-neighbor graph. For centered atom coordinates $\rho_i\in\mathbb{R}^3$, node features are
\begin{equation}
  \begin{aligned}
  s_i & = [\operatorname{onehot}(a_i);0]\in\mathbb{R}^{11},\\
  V_i & = [\rho_i,\rho_i/\|\rho_i\|_2,\mathbf{0}]
  \in\mathbb{R}^{3\times 3},
  \end{aligned}
\end{equation}
where $a_i$ is the atom element. For each directed edge $(i,j)$, edge features are the unit direction $u_{ij}=(\rho_j-\rho_i)/\|\rho_j-\rho_i\|_2$ and a 32-dimensional scalar vector consisting of 16 Gaussian distance features with centers on $[0,8]$ and width 0.5, concatenated with a fixed edge-type indicator. A three-layer geometric vector perceptron encoder maps the graph to node states, and the pocket vector is their mean pooled scalar output,
\begin{equation}
  g_{\mathrm{pocket}} = \frac{1}{|\mathcal{V}|}\sum_{i\in \mathcal{V}} W_{\mathrm{out}}\bigl(h_i^{(3)}\bigr) \in \mathbb{R}^{256}.
\end{equation}
This vector is projected by the pocket-condition projector and injected at the pocket-condition value token, keeping the molecular decoder unchanged while allowing the model to use three-dimensional pocket context.

\subsection{Training Data}

The base pretraining and molecule-context SFT corpus is derived from UniChem \cite{Chambers2013UniChem}. We canonicalize molecules as non-isomeric RDKit SMILES with explicit hydrogens removed and stereochemistry stripped, deduplicate them, and filter out malformed records, mixtures or reactions, RDKit sanitization failures, molecules with more than 50 heavy atoms, radicals, molecular weight above 800\,Da, QED below 0.3 \cite{Bickerton2012QED}, SAS above 5 \cite{Ertl2009SAScore}, and PAINS \cite{Baell2010PAINS}, Brenk \cite{Brenk2008ScreeningLibraries}, or NIH \cite{Jadhav2010Artifacts,Doveston2015LeadOriented} structural alerts. The final UniChem-derived molecule list, with SAIR and PLINDER training-set ligand coverage included, is converted offline into randomized \fselfies{} strings. The full molecule-context file contains 135{,}763{,}524 \fselfies{} records, corresponding to 33{,}940{,}881 molecule-condition rows with four precomputed randomized folds. The aligned condition files contain the nine scalar molecular properties and a 2D pharmacophore fingerprint.

Target-context SFT uses protein-ligand pairs from SAIR and the PLINDER training set \cite{Lemos2025SAIR,Durairaj2024PLINDER}. Ligands are normalized to no-hydrogen, no-stereochemistry SMILES, filtered and deduplicated, and converted to 10-fold randomized \fselfies{} files. Protein-sequence context is stored as precomputed ESMC-600M embeddings, and pocket context is read from processed no-hydrogen pocket structures. To control leakage in CrossDocked2020 evaluation \cite{Francoeur2020CrossDocked}, we removed SAIR and PLINDER training-set pairs whose protein sequence had more than 30\% identity to any CrossDocked2020 test protein, using MMseqs2 sequence search \cite{Steinegger2017MMseqs2,Kallenborn2025GPUMMseqs2}. After this filtering, the target-context pool contains 573{,}463 SAIR pair records and 21{,}770 PLINDER training-set pair records.

\subsection{Training Objectives}

\molexar{} uses a two-stage recipe. In the first stage, the base model is pretrained on randomized \fselfies{} strings using causal language modeling. The template includes the condition block, but no condition values are injected and the condition encoders and GVP pocket encoder are not updated by this objective. This stage teaches the decoder the \fselfies{} syntax, fragment-tree traversal patterns, and molecular distribution. In the second stage, universal multi-condition SFT starts from the pretrained checkpoint and trains the decoder, condition encoders, and pocket encoder on condition-molecule pairs.

For both stages, the model predicts the next token from the shifted logits. If $h_t=F_\theta(\tilde{e}_{0:t})$ is the decoder state, the vocabulary logits use the implementation's final softcap
\begin{equation}
  o_t = \tau_{\mathrm{out}}\tanh\!\left(\frac{W_{\mathrm{lm}}h_t}{\tau_{\mathrm{out}}}\right),
  \qquad \tau_{\mathrm{out}}=30.
\end{equation}
The training loss is
\begin{equation}
  \mathcal{L}(\theta) =
  -\frac{1}{\sum_{b,t} m_{bt}}
  \sum_b \sum_{t=1}^{T-1} m_{bt}
  \log \operatorname{softmax}(o_{b,t-1})_{x_{bt}},
\end{equation}
where $m_{bt}=1$ when the label at position $t$ is not ignored. Stage 1 uses the full shifted template labels. Stage 2 sets $m_{bt}=0$ for the prefix through \texttt{<MOL>} and for padding tokens, so the loss is applied to the molecular continuation and closing tokens conditioned on the prompt and injected values.

Universal SFT mixes molecule-context and target-context samples in a 4:1 ratio. For molecule-context samples, conditions are drawn from the nine scalar properties and the 2D pharmacophore fingerprint. The default schedule samples one, two, or three active molecule-side conditions with probabilities 0.6, 0.3, and 0.1, respectively, and oversamples the pharmacophore-fingerprint condition with probability 0.5. Thus molecule-context SFT covers up to 175 single-, pair-, and triple-condition subsets; the property-only subsets account for 129 of these combinations. For target-context samples, sequence and pocket conditions are sampled first: when both are available, the dual sequence-plus-pocket condition is selected with probability 0.2 and otherwise one of the two target modalities is sampled. Ligand-side property or pharmacophore conditions can then be added with probability 0.2, capped at three active condition values in total. Both stages are trained with maximum sequence length 256, learning rate $2\times10^{-4}$, 2{,}000 warmup steps, batch size 1{,}000, bfloat16 mixed precision, and full-shard FSDP on eight H800 GPUs; Stage 1 uses two epochs and Stage 2 uses five epochs.

\section{Experiments}

\subsection{Unconditional and Fragment-Constrained Generation}

\subsubsection{Unconditional Generation}

We first evaluate unconditional generation, which probes the ability of the base model to produce valid molecules that span a broad region of chemical space. We sampled 10,000 molecules from our pretrained base model \molexar{} and assessed them with four metrics: validity, uniqueness, diversity, and quality. Validity is the fraction of generated strings that are parseable by RDKit; uniqueness is the fraction of distinct molecules among the valid ones; and diversity is the mean pairwise Tanimoto distance over ECFP4 fingerprints of the valid molecules. Quality is the fraction of molecules that are simultaneously drug-like and synthesizable, defined as drug-likeness $\text{QED} \geq 0.6$ \cite{Bickerton2012QED} and synthetic accessibility $\text{SAS} \leq 4$ \cite{Ertl2009SAScore}, following \citet{Jin2020MultiObj} and \citet{Lee2025GenMol}. We compare against SAFE-GPT \cite{Noutahi2024SAFE}, GenMol \cite{Lee2025GenMol}, and ChemFM-1B and ChemFM-3B \cite{Cai2025ChemFM}, all of which are language models trained on molecular string representations.

\begin{table}[t]
  \caption{\textbf{Unconditional molecule generation results on 10,000 samples.} \textbf{Bold}: best; \underline{underline}: second best.}
  \label{tab:unconditional}
  \begin{center}
  \footnotesize
  \setlength{\tabcolsep}{3pt}
  \begin{tabular}{l c c c c}
    \toprule
    Method    & Validity  & Uniqueness  & Diversity  & Quality  \\
    \midrule
    SAFE-GPT  & 0.9351             & \textbf{1.0000}    & 0.8783             & 0.5465             \\
    GenMol    & \underline{0.9997} & 0.8132             & 0.8324             & \underline{0.7428} \\
    ChemFM-1B & 0.9939             & \textbf{1.0000}    & \textbf{0.9001}    & 0.4157             \\
    ChemFM-3B & 0.9945             & \textbf{1.0000}    & \textbf{0.9001}    & 0.4090             \\
    \midrule
    \molexar{} & \textbf{1.0000}   & \underline{0.9997} & \underline{0.8824} & \textbf{0.8326}    \\
    \bottomrule
  \end{tabular}
  \end{center}
  \vskip -0.1in
\end{table}

\begin{figure}[t]
  \begin{center}
    \includegraphics[width=\columnwidth]{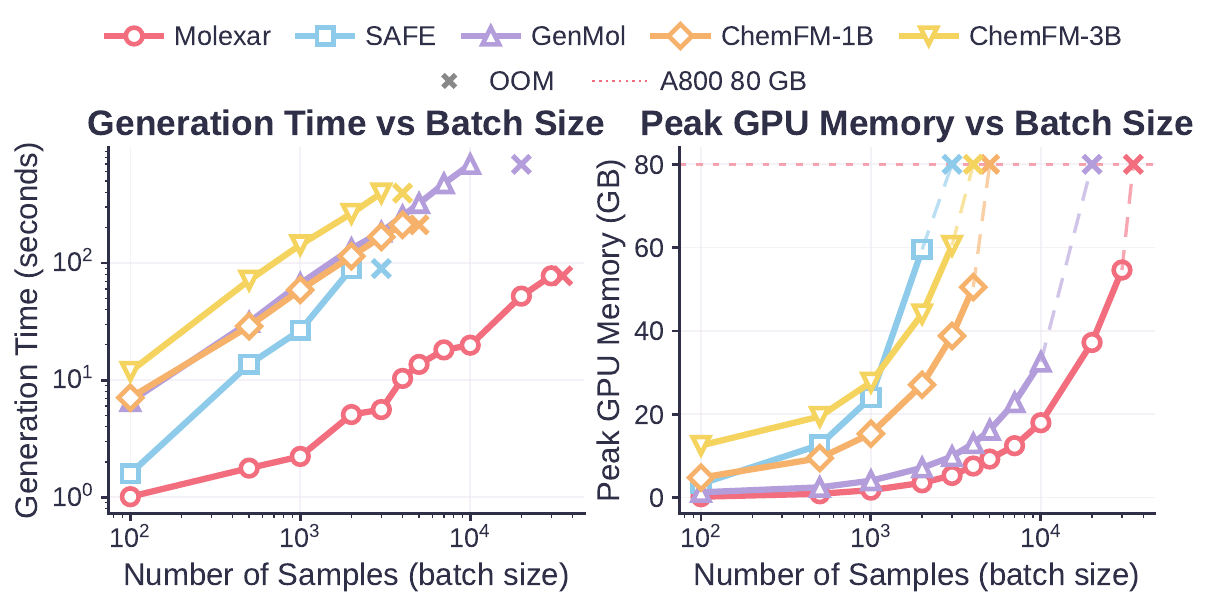}
    \caption{\textbf{Inference efficiency under unconditional generation.}
      Generation time (left) and peak GPU memory (right) versus batch size. Crosses ($\times$): first OOM
      batch size; dotted line: 80\,GB capacity.}
    \label{fig:efficiency}
  \end{center}
  \vskip -0.1in
\end{figure}

As shown in \cref{tab:unconditional}, \molexar{} attains the best overall profile, combining the highest quality with strong diversity and near-perfect uniqueness. By construction, \molexar{} decodes generated \fselfies{} strings through a validity-oriented codec with RDKit sanitization and non-strict recovery paths, yielding $100\%$ validity in our samples. This validity profile holds across all subsequent experiments and illustrates a concrete advantage of \fselfies{} over SAFE \cite{Noutahi2024SAFE} and SMILES \cite{Weininger1988SMILES}. The agreement between the sampled molecules and the training distribution is examined in \cref{sec:distribution-overlap}. Example generated molecules are shown in \cref{fig:unconditional-mols}.

\begin{table*}[b]
\caption{\textbf{Fragment-constrained molecule generation results.} The results are the means and the standard deviations of three runs. The best results are highlighted in \textbf{bold}.}
\label{tab:fragment-constrained}
\begin{center}
\footnotesize
\begin{tabular}{ll cc c cc}
\toprule
Method & Task & Validity (\%) & Uniqueness (\%) & Quality (\%) & Diversity & Distance \\
\midrule
\multirow{5}{*}{SAFE-GPT}
 & Linker design            & 46.7 $\pm$ 1.7 & \textbf{72.5 $\pm$ 8.7} & 15.6 $\pm$ 1.7 & 0.497 $\pm$ 0.024 & 0.509 $\pm$ 0.033 \\
 & Scaffold morphing        & 49.2 $\pm$ 2.4 & \textbf{76.5 $\pm$ 1.1} & 13.7 $\pm$ 1.4 & 0.513 $\pm$ 0.005 & 0.540 $\pm$ 0.002 \\
 & Motif extension          & 91.5 $\pm$ 6.4 & 69.9 $\pm$ 3.1 & 19.7 $\pm$ 3.3 & 0.542 $\pm$ 0.010 & 0.670 $\pm$ 0.006 \\
 & Scaffold decoration      & 81.6 $\pm$ 6.9 & 79.0 $\pm$ 3.2 & 9.4 $\pm$ 2.7  & 0.576 $\pm$ 0.008 & 0.617 $\pm$ 0.005 \\
 & Superstructure generation& 97.4 $\pm$ 0.4 & 86.1 $\pm$ 3.7 & 18.0 $\pm$ 3.9 & 0.573 $\pm$ 0.021 & \textbf{0.785 $\pm$ 0.026} \\
\midrule
\multirow{5}{*}{GenMol}
 & Linker design            & 93.9 $\pm$ 2.9 & 71.7 $\pm$ 1.1 & 13.9 $\pm$ 0.4 & \textbf{0.567 $\pm$ 0.001} & \textbf{0.555 $\pm$ 0.002} \\
 & Scaffold morphing        & 93.9 $\pm$ 2.9 & 71.7 $\pm$ 1.1 & 13.9 $\pm$ 0.4 & \textbf{0.567 $\pm$ 0.001} & \textbf{0.555 $\pm$ 0.002} \\
 & Motif extension          & 82.2 $\pm$ 0.1 & 57.6 $\pm$ 1.9 & 16.5 $\pm$ 0.5 & \textbf{0.630 $\pm$ 0.002} & \textbf{0.677 $\pm$ 0.003} \\
 & Scaffold decoration      & 98.0 $\pm$ 0.2 & 86.1 $\pm$ 1.1 & \textbf{32.5 $\pm$ 1.1} & \textbf{0.598 $\pm$ 0.002} & \textbf{0.626 $\pm$ 0.002} \\
 & Superstructure generation& 98.2 $\pm$ 0.5 & \textbf{87.6 $\pm$ 4.1} & 29.5 $\pm$ 2.8 & \textbf{0.613 $\pm$ 0.024} & 0.757 $\pm$ 0.015 \\
\midrule
\multirow{5}{*}{Molexar}
 & Linker design            & \textbf{100.0 $\pm$ 0.0} & 55.1 $\pm$ 0.3 & \textbf{25.2 $\pm$ 1.7} & 0.536 $\pm$ 0.004 & 0.529 $\pm$ 0.002 \\
 & Scaffold morphing        & \textbf{100.0 $\pm$ 0.0} & 55.1 $\pm$ 0.3 & \textbf{25.2 $\pm$ 1.7} & 0.536 $\pm$ 0.004 & 0.529 $\pm$ 0.002 \\
 & Motif extension          & \textbf{100.0 $\pm$ 0.0} & \textbf{79.3 $\pm$ 1.8} & \textbf{46.8 $\pm$ 1.6} & 0.621 $\pm$ 0.004 & 0.652 $\pm$ 0.000 \\
 & Scaffold decoration      & \textbf{100.0 $\pm$ 0.0} & \textbf{96.3 $\pm$ 0.5} & 16.6 $\pm$ 0.4 & 0.580 $\pm$ 0.005 & 0.601 $\pm$ 0.003 \\
 & Superstructure generation& \textbf{100.0 $\pm$ 0.0} & 70.5 $\pm$ 5.8 & \textbf{31.9 $\pm$ 0.3} & 0.487 $\pm$ 0.016 & 0.746 $\pm$ 0.016 \\
\bottomrule
\end{tabular}
\end{center}
\vskip -0.1in
\end{table*}

We further benchmark inference latency and peak GPU memory for all methods on a single NVIDIA A800-SXM4-80GB GPU in FP32 (\cref{fig:efficiency}). \molexar{} attains the lowest latency and memory footprint across all batch sizes, scaling to far larger batches before saturating the GPU. It samples up to 30,000 molecules within a single batch in under 80\,s, whereas the baselines typically run out of memory once the batch size exceeds 10,000. This efficiency follows from \molexar{}'s small parameter count and the Gemma2 backbone \cite{GemmaTeam2024Gemma2}.

\subsubsection{Fragment-Constrained Generation}

Fragment-constrained generation grows new fragments onto one or more user-specified input fragments. Using the benchmark of \citet{Noutahi2024SAFE}, which provides ten known drugs together with their fragment decompositions, we evaluate five tasks: linker design, scaffold morphing, motif extension, scaffold decoration, and superstructure generation. We characterize the generated molecules with validity, uniqueness, diversity, and quality, as well as distance, the average Tanimoto distance between the original and the generated molecules. For each drug and each task we generated 100 molecules and compared against SAFE-GPT \cite{Noutahi2024SAFE} and GenMol \cite{Lee2025GenMol}. Note that linker design and scaffold morphing use the same generation procedure for both \molexar{} and GenMol.

As shown in \cref{tab:fragment-constrained}, \molexar{} outperforms both baselines in validity on every task and in quality on most tasks, indicating that it produces more drug-like molecules and is well suited to realistic drug-design settings. Examples of \molexar{} generations for Eliglustat are shown in \cref{fig:frag-constrained}.

\subsection{Property-Controlled Generation}

Conditioning in prior molecular generators is often restricted: some methods require additional, goal-specific reinforcement learning to steer generation toward a target property
 \cite{Noutahi2024SAFE}, while others condition on only a small number of continuous properties \cite{Cai2025ChemFM}. Through supervised fine-tuning (SFT) against a universal objective,
  \molexar{} instead conditions generation on nine continuous or discrete properties and supports single-, dual-, and triple-property prompts. We note that the model is explicitly trained on these combinations: the SFT schedule samples all 129 property-only subsets (9 singletons, 36 pairs, and 84 triples).

\subsubsection{Single-Property Conditioning}

\begin{figure}[t]
  \begin{center}
    \centerline{\includegraphics[width=0.95\columnwidth]{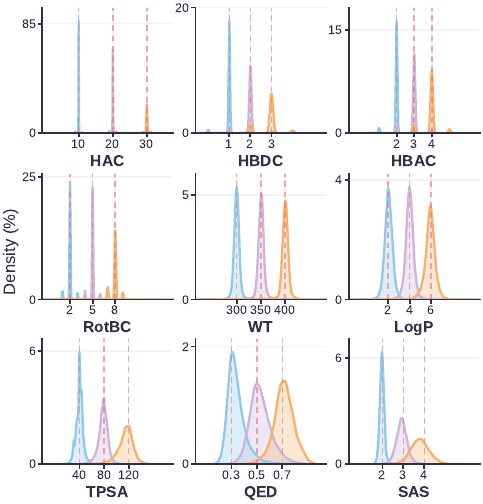}}
    \caption{\textbf{Single-property conditioning across nine properties.} Each panel
      shows the property distribution of molecules generated under three
      target values (dashed red lines); from left to right and top to bottom:
      HAC, HBDC, HBAC, RotBC (discrete), WT, LogP, TPSA, QED, SAS
      (continuous).}
    \label{fig:single-property}
  \end{center}
  \vskip -0.3in
\end{figure}

For each of the nine properties we selected three target values and generated 10,000 molecules per target, as shown in \cref{fig:single-property}. \molexar{} exhibits
 strong conditioning ability for both discrete and continuous properties: the generated molecules form a single sharp peak around each target value, with the mode of
  the distribution coinciding with the requested target, demonstrating reliable single-property instruction following. 

We note that the degree of single-property obedience is influenced by the abundance of the corresponding property values in the training set; \cref{sec:obedience-rate} reports the relationship between the obedience rate and training-set abundance.

\subsubsection{Multi-Property Conditioning}

\molexar{} also supports conditioning on combinations of properties. This is important because many drug-design objectives are defined by several properties jointly rather than in isolation, as in the rule of three for fragment-based discovery, Lipinski's rule of five, and the beyond-rule-of-five (bRo5) regime for oral drugs \cite{Congreve2003RuleOfThree,Lipinski2001Ro5,Doak2014bRo5}. Because the number of possible combinations is large, we evaluate three representative scenarios, shown in \cref{fig:multi-property}.

The first scenario fixes HAC and varies RotBC (\cref{fig:multi-property}, left), testing whether the model can produce molecules of different flexibility at a constant size. The generated molecules follow the requested targets closely, and they shift visibly from fused-ring scaffolds toward longer flexible chains, indicating that \molexar{} captures the joint distribution rather than each property in isolation.

\begin{strip}
  \centering
  \includegraphics[width=\textwidth]{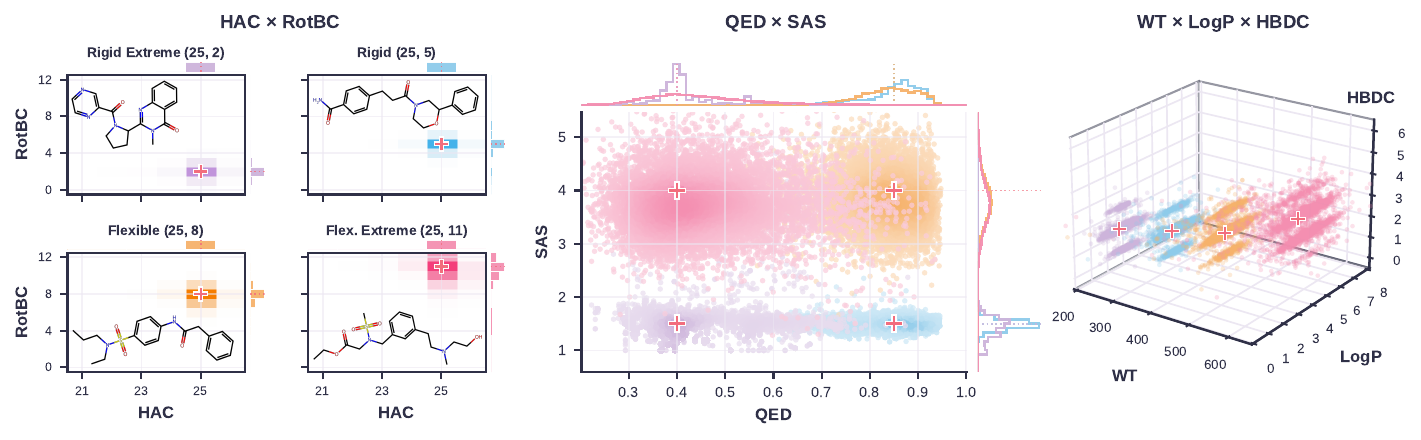}
  \captionof{figure}{\textbf{Multi-property conditioning under realistic design scenarios.}
    \textbf{Left:} size-flexibility decoupling, jointly conditioning on
    \{HAC, RotBC\} at rigid-extreme \{25, 2\}, rigid \{25, 5\}, flexible
    \{25, 8\}, and flexible-extreme \{25, 11\}.
    \textbf{Middle:} QED-SAS trade-off, conditioning on \{QED, SAS\} at
    \{0.85, 1.5\}, \{0.85, 4.0\}, \{0.40, 1.5\}, and \{0.40, 4.0\}.
    \textbf{Right:} drug-development regimes defined by \{WT, LogP, HBDC\}:
    fragment-like \{250, 1.5, 2\}, lead-like \{350, 2.5, 2\}, drug-like
    \{450, 3.5, 2\}, and beyond-rule-of-five (bRo5) \{600, 4.5, 3\}.}
  \label{fig:multi-property}
\end{strip}

\begin{figure*}[!t]
  \begin{center}
    \centerline{\includegraphics[width=0.8\textwidth]{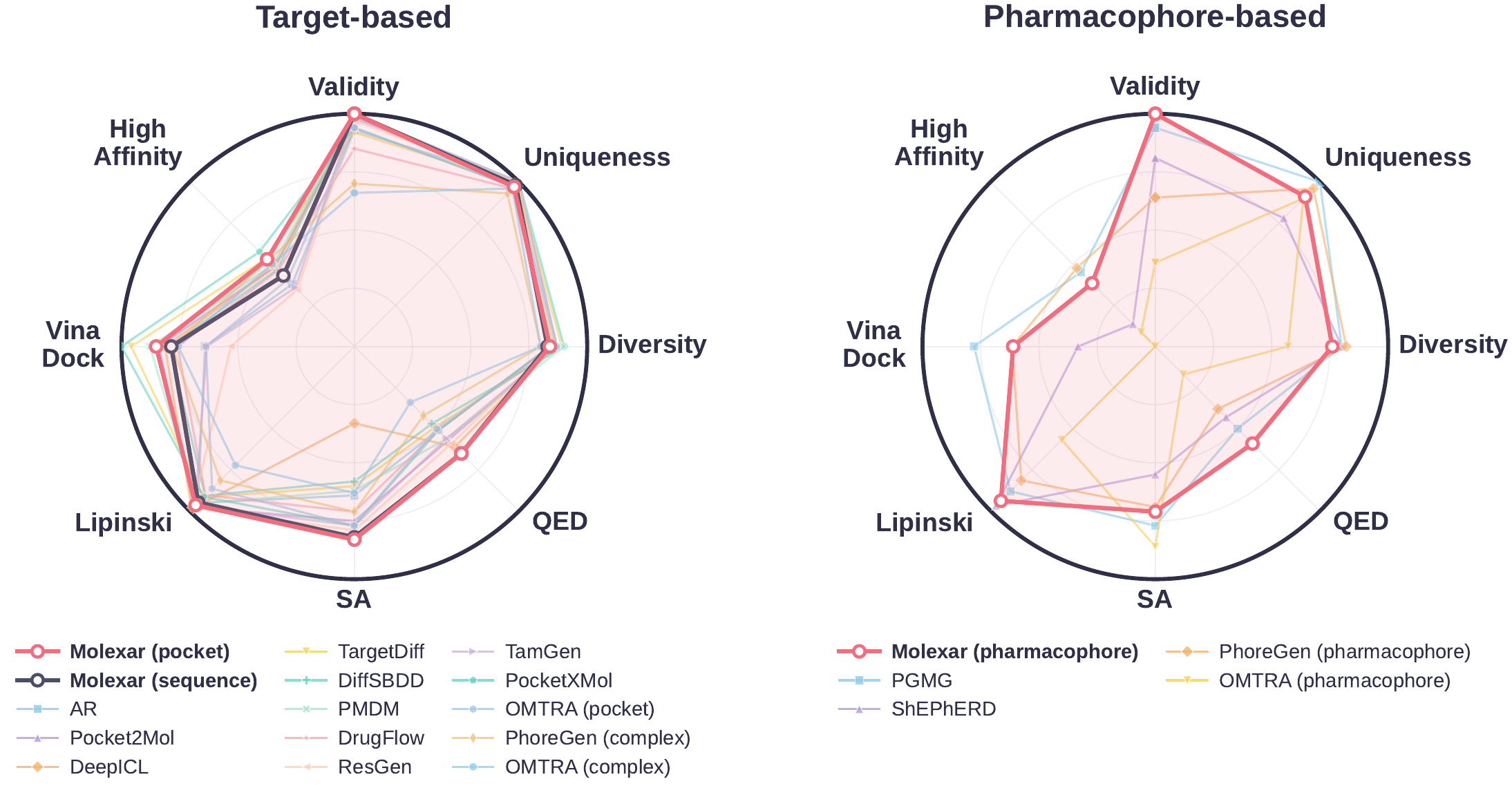}}
    \caption{\textbf{Target-conditioned molecular generation on the CrossDocked2020 test
      set.} \textbf{Left:} generation metrics for methods conditioned on target
      structure or sequence (some methods additionally take a reference-ligand
      pharmacophore as input). \textbf{Right:} generation conditioned only on
      the reference-ligand pharmacophore. For \molexar{}, we report molecules
      generated under each of three conditioning modes: protein sequence,
      binding pocket, and ligand pharmacophore.}
    \label{fig:target-radar}
  \end{center}
  \vskip -0.4in
\end{figure*}

The second scenario combines QED and SAS (\cref{fig:multi-property}, middle) to test whether the model can produce molecules that are both drug-like and easy to synthesize. We sampled the four corners of the QED-SAS plane, including settings in which one or both targets are placed at unfavorable values. The model follows all four settings, and obedience is highest in the most useful corner, namely high QED together with low SAS. This suggests that the training corpus is well matched to drug-like chemistry.

The third scenario uses three properties, WT, LogP, and HBDC, to trace a rule-of-five drug-likeness trajectory, simulating the shift in chemical space along a drug-discovery pipeline from fragment-like to bRo5 regimes (\cref{fig:multi-property}, right). \molexar{} tracks the requested WT, LogP, and HBDC closely across the four regimes, and the generated molecules respect the rule-of-three, rule-of-five, and bRo5 criteria that define them, demonstrating its drug-generation capability. The bRo5 regime is matched slightly less tightly than the others, consistent with a training corpus filtered to drug-like molecules (see Methods). This experiment is a proof of concept that \molexar{} can populate molecular libraries spanning different stages of drug development. The representative molecules are shown in \cref{fig:drug-mol}.

We note that the conditioned properties are not independent. Correlations among them mean that some combinations correspond to molecules that are scarce or absent in the training set, which can reduce obedience for those settings. We report the pairwise property correlations of the training set in \cref{sec:training-property-correlation} for reference.

\subsection{Target-Conditioned Generation}

\begin{figure*}[!t]
  \begin{center}
    \centerline{\includegraphics[width=\textwidth]{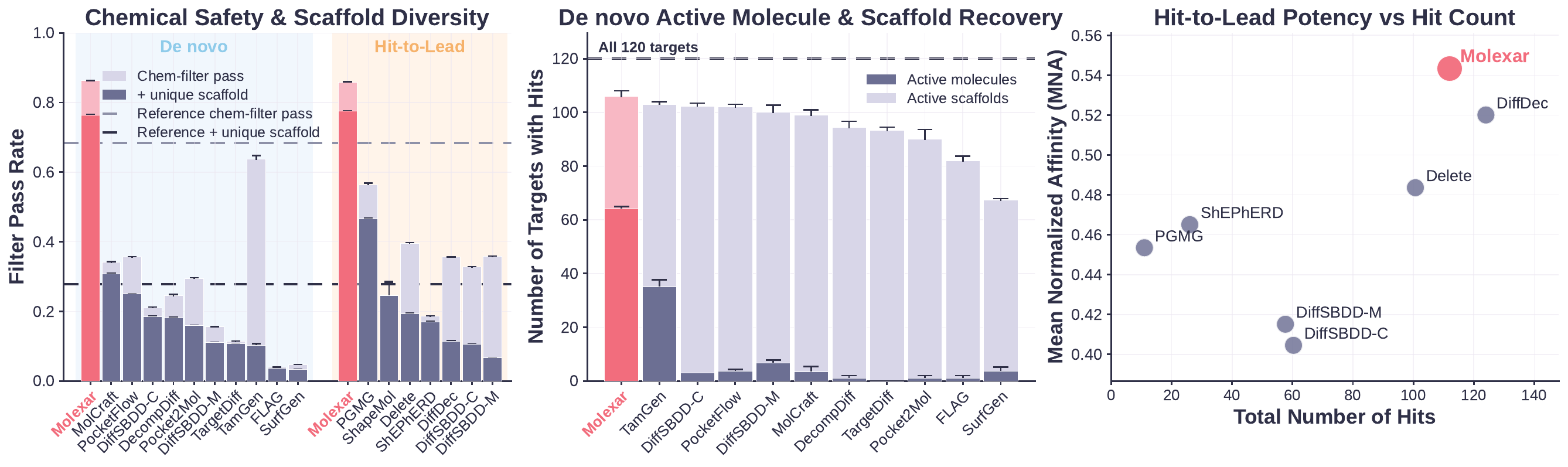}}
    \caption{\textbf{Real-world applicability results on MolGenBench.}
      \textbf{Left:} fraction of generated molecules that pass the chemical
      filters (dark) and the fraction of molecules that also contribute a
      unique Bemis-Murcko scaffold \cite{Bemis1996Frameworks} (light).
      \textbf{Middle:} evaluation of molecular generative models on identifying
      active molecules and scaffolds through \textit{de novo} generation.
      \textbf{Right:} evaluation of hit-to-lead (H2L) optimization, measured by
      the potency of generated active molecules and the number of hits.}
    \label{fig:molgenbench-main}
  \end{center}
  \vskip -0.1in
\end{figure*}

To assess whether \molexar{} is target-aware, we evaluate it on the CrossDocked2020 test set \cite{Francoeur2020CrossDocked}. The baselines fall into three groups: pocket-only methods (AR \cite{Luo2022SBDD}, Pocket2Mol \cite{Peng2022Pocket2Mol}, DeepICL \cite{Zhung2024DeepICL}, TargetDiff \cite{Guan2023TargetDiff}, DiffSBDD \cite{Schneuing2024DiffSBDD}, PMDM \cite{Huang2024PMDM}, DrugFlow \cite{Schneuing2025DrugFlow}, ResGen \cite{Zhang2023ResGen}, TamGen \cite{Wu2024TamGen}, PocketXMol \cite{Peng2026PocketXMol}, and OMTRA \cite{Dunn2025OMTRA} (pocket)); pharmacophore-only methods that condition on a reference-ligand pharmacophore (PGMG \cite{Zhu2023PGMG}, ShEPhERD \cite{Adams2025ShEPhERD}, PhoreGen \cite{Peng2025PhoreGen} (pharmacophore), and OMTRA (pharmacophore)); and methods that condition jointly on the pocket and the pharmacophore (PhoreGen (complex) and OMTRA (complex)). Some methods support more than one of these settings (PhoreGen and OMTRA). Owing to its flexible conditioning interface, \molexar{} is evaluated under three settings: pocket geometry (\molexar{} (pocket)), protein sequence (\molexar{} (sequence)), and reference-ligand pharmacophore (\molexar{} (pharmacophore)). To prevent data leakage, we removed training samples with high sequence homology ($>30\%$) to the CrossDocked2020 test set. For each of the 100 complexes in the test set we generated 100 molecules and report validity, uniqueness, diversity, QED, SA (rescaled from the original 1--10 scale to 0--1, where higher is better), Lipinski's rule of five, the Vina docking score \cite{Trott2010Vina}, and the fraction of molecules whose Vina docking score is better than that of the reference ligand (high affinity).

As shown in \cref{fig:target-radar}, \molexar{} achieves a well-balanced profile across all metrics. Under every conditioning setting, the molecules generated by \molexar{} rank among the best in SA and QED, indicating favorable drug-likeness. Although it is a language model, \molexar{} (pocket) attains better Vina docking scores than many structure-based models and clearly outperforms the language-model baseline TamGen, ranking third in high-affinity ratio among the target-based methods. Despite the strict homology-based deduplication, \molexar{} (sequence) still achieves strong Vina scores, reflecting the out-of-distribution generalization afforded by multi-condition SFT. For pharmacophore conditioning, \molexar{} performs comparably to existing models. Detailed per-metric results on CrossDocked2020 are reported in \cref{tab:target,tab:pharmacophore}. Overall, \molexar{} demonstrates a balanced and strong ability to generate potentially high-affinity molecules for specific targets.

\subsection{Real-World Applicability on MolGenBench}

To evaluate the practical applicability of \molexar{}, we test it on MolGenBench \cite{Cao2025MolGenBench}. The benchmark comprises two tasks, \textit{de novo} generation and hit-to-lead (H2L) optimization, and assesses both generation across 120 protein targets and the recovery of ground-truth drug molecules. We follow its protocol strictly: for \textit{de novo} generation we condition on the pocket, and for the H2L task we condition jointly on the pocket and the reference-ligand pharmacophore.

We focus on the practical drug-likeness of the generated molecules. Viable drugs must be sufficiently safe and free of unstable or reactive groups, and MolGenBench applies a panel of industry-standard chemical filters \cite{Schuffenhauer2020ScreeningDeck,Bruns2012ReactiveCompounds}; under these filters, only about 70\% of the reference active molecules and about 30\% of their scaffolds pass. As shown in \cref{fig:molgenbench-main} (left), \molexar{} reaches a molecule pass rate of about 85\% and a scaffold pass rate of about 75\%, indicating high safety and diversity. For the \textit{de novo} task (\cref{fig:molgenbench-main}, middle), \molexar{} exactly recovers the active-molecule SMILES for more than half of the targets and nearly all of the active scaffolds. For the H2L task (\cref{fig:molgenbench-main}, right), \molexar{} discovers the second-largest number of active molecules while achieving the highest potency, demonstrating its ability to generate both a greater quantity of active molecules and compounds with superior bioactivity. Additionally, results on general molecular properties are reported in \cref{fig:molgenbench-properties}, where \molexar{} ranks among the top methods. These results indicate that \molexar{} is an effective model for generating safe, high-affinity molecules in realistic drug-discovery settings.

\section{Conclusions}

In this work, we present \molexar{}, a unified molecular foundation model that treats diverse design constraints as values in a single autoregressive prompt. \molexar{} is built on two components: \fselfies{}, an explicit BRICS-fragment molecular language with validity-preserving decoding and fragment continuation, and a value-token embedding replacement mechanism that binds scalar properties, pharmacophore fingerprints, protein-sequence embeddings, and pocket-geometry embeddings to a single Gemma2-style decoder. This removes the need for task-specific generators and cross-attention, letting unconditional sampling, property control, pharmacophore guidance, target and pocket conditioning, and fragment-constrained generation share one sequence template and a cache-compatible decoding path.

This unified interface comes at no cost to generation quality. In unconditional generation, \molexar{} attains 100\% validity, near-perfect uniqueness, high diversity, and the best quality score among molecular-language baselines, all at the lowest latency and memory use; it preserves perfect validity under fragment constraints and follows single-, dual-, and triple-property instructions from fragment-like to bRo5 regimes. Conditioned on protein sequence, pocket geometry, or pharmacophores on CrossDocked2020, it generates chemically favorable molecules with competitive docking metrics under leakage-controlled training, and on MolGenBench it combines high chemical-filter pass rates with strong active-molecule and scaffold recovery and H2L potency. Together, these results establish \molexar{} as a compact, efficient foundation model for multimodal molecular design, and future work should test prospective, experimentally validated design cycles.

\section*{Impact Statement}


This paper presents \molexar{}, a unified molecular foundation model whose goal is to advance molecule generation. By handling diverse design conditions within a single compact and fast model that produces drug-like, target-aware candidates, \molexar{} lowers the cost of exploring chemical space and holds broad potential for drug discovery.

\section*{Code Availability}

The source code for \molexar{} is publicly available at \url{https://github.com/fairydance/Molexar}, and the implementation of \fselfies{} is available at \url{https://github.com/fairydance/Fragment-SELFIES}. Additional information and resources for both projects are available at \url{https://molexar.com}.

\section*{Acknowledgements}

This work was supported in part by the National Natural Science Foundation of China (grant nos. 22033001 and T2321001), the National Key R\&D Program of China (grant no. 2023YFF1205103), the Chinese Academy of Medical Sciences (grant no. 2021-I2M-5-014), and Anhui's Plans for Major Provincial Science \& Technology Projects (grant no. 202303a07020009). We thank the team managing the high-performance computing platform at the Peking-Tsinghua Center for Life Sciences, Peking University, for providing computational resources.

\bibliography{molexar_manuscript}
\bibliographystyle{icml2026}

\newpage
\appendix
\onecolumn



\section{Alignment of Unconditionally Generated Molecules with the Training Distribution}
\label{sec:distribution-overlap}

\begin{figure}[h]
  \begin{center}
    \includegraphics[width=\columnwidth]{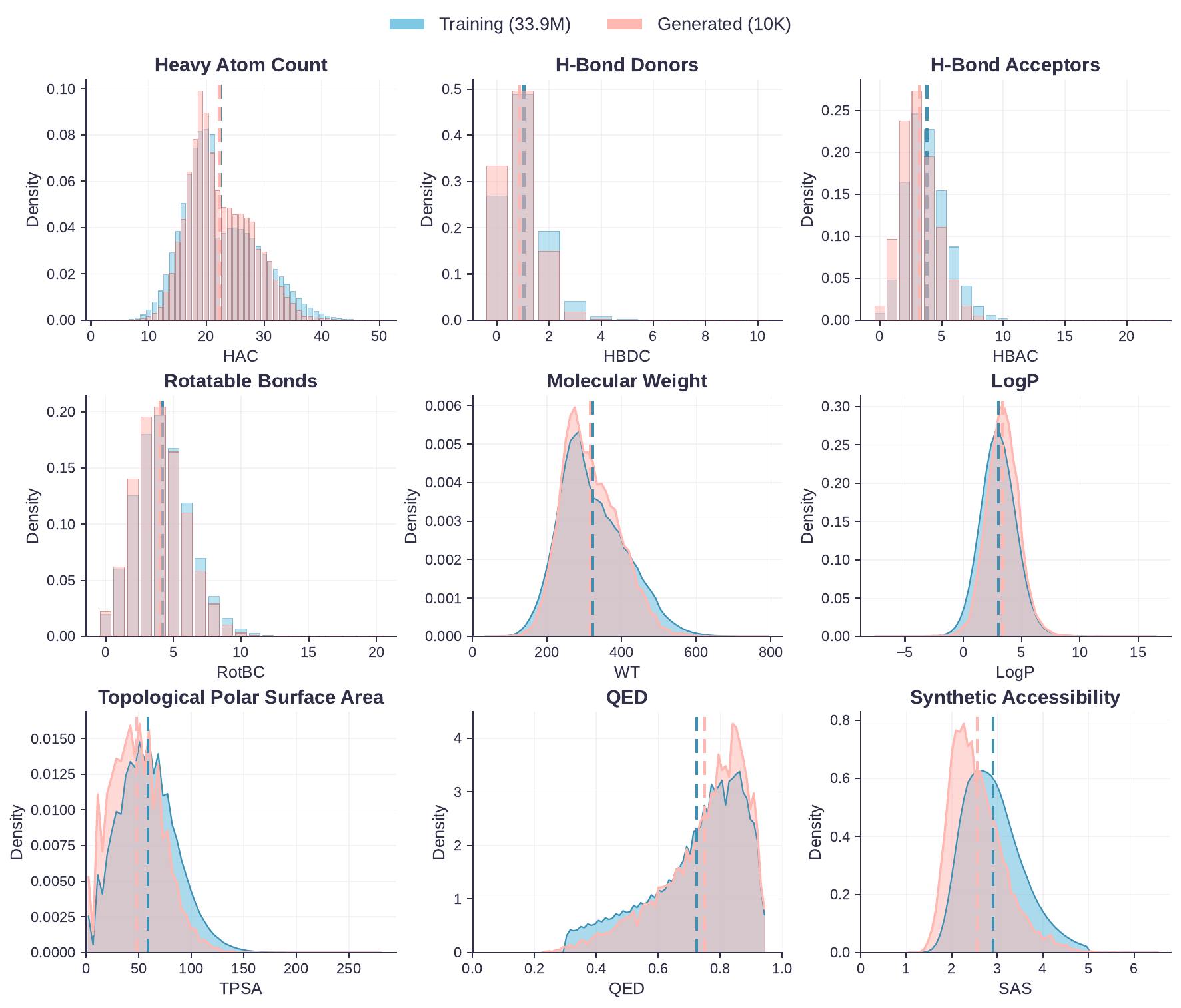}
  \end{center}
  \caption{
    \textbf{Marginal property distributions of unconditionally generated molecules
    versus the training set.} Each panel overlays the normalized histogram
    and kernel-density estimate of one molecular property for 10,000
    molecules sampled from \molexar{} (pink) and for the 33.9M-molecule
    training corpus (green); dashed vertical lines mark the respective
    means.
  }
  \label{fig:property-overlay}
  \vskip -0.1in
\end{figure}

\begin{figure}[t]
  \begin{center}
    \includegraphics[width=0.8\columnwidth]{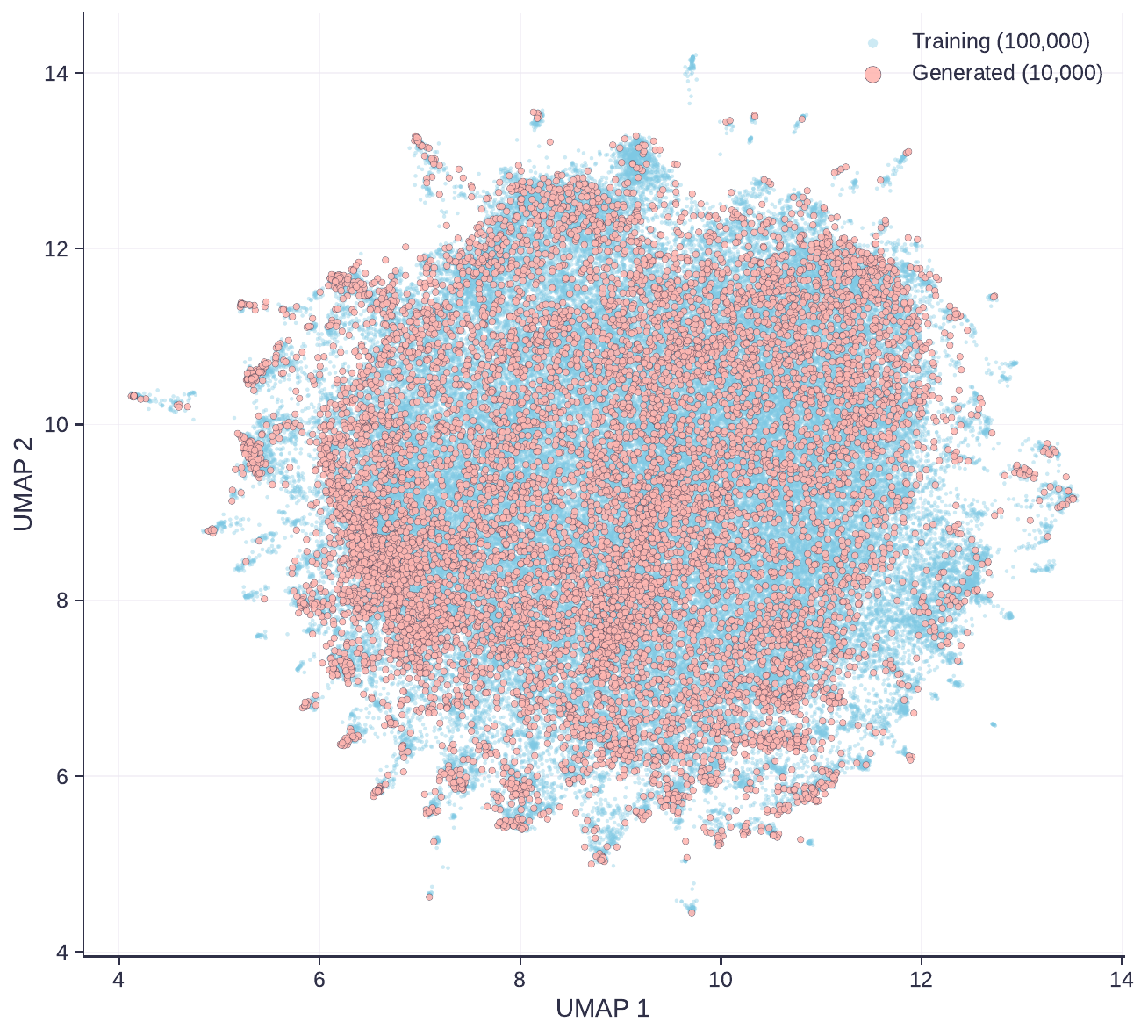}
  \end{center}
  \caption{
    \textbf{UMAP embedding of ECFP4 fingerprints for 10,000 unconditionally
    generated molecules (pink) and 100,000 molecules subsampled from the
    training set (green).}
  }
  \label{fig:umap}
  \vskip -0.1in
\end{figure}

To verify that unconditional sampling reproduces the chemistry of the pretraining corpus rather than collapsing onto a narrow region or drifting into unrealistic chemical space, we compare 10,000 molecules sampled from \molexar{} against the training set along two complementary views. \Cref{fig:property-overlay} overlays the marginal distributions of nine molecular properties, and \cref{fig:umap} compares the two populations in a shared low-dimensional embedding of their structural fingerprints. Across both views the generated molecules closely track the training distribution, indicating faithful distribution learning without mode collapse.

\clearpage

\section{Single-Property Obedience versus Training-Set Abundance}
\label{sec:obedience-rate}

To quantify how reliably single-property conditioning obeys the requested instruction, and in particular whether obedience is retained for target values that are scarce in the training set, we conducted a detailed analysis of the per-property obedience rate as a function of training-set abundance. For discrete properties, we enumerated all available sampling points and defined the obedience rate as the fraction of generated molecules whose property value exactly matches the requested point. For continuous properties, we spaced the sampling points by the RBF encoding step size and defined the obedience rate as the fraction of molecules falling within a tolerance window of the target, where the tolerance is set to that same step size. The sampling ranges, step counts, and tolerances for all nine properties are listed in \cref{tab:property-config}, and we generated 1{,}000 molecules at each sampling point.

\begin{figure}[h]
  \begin{center}
    \includegraphics[width=\columnwidth]{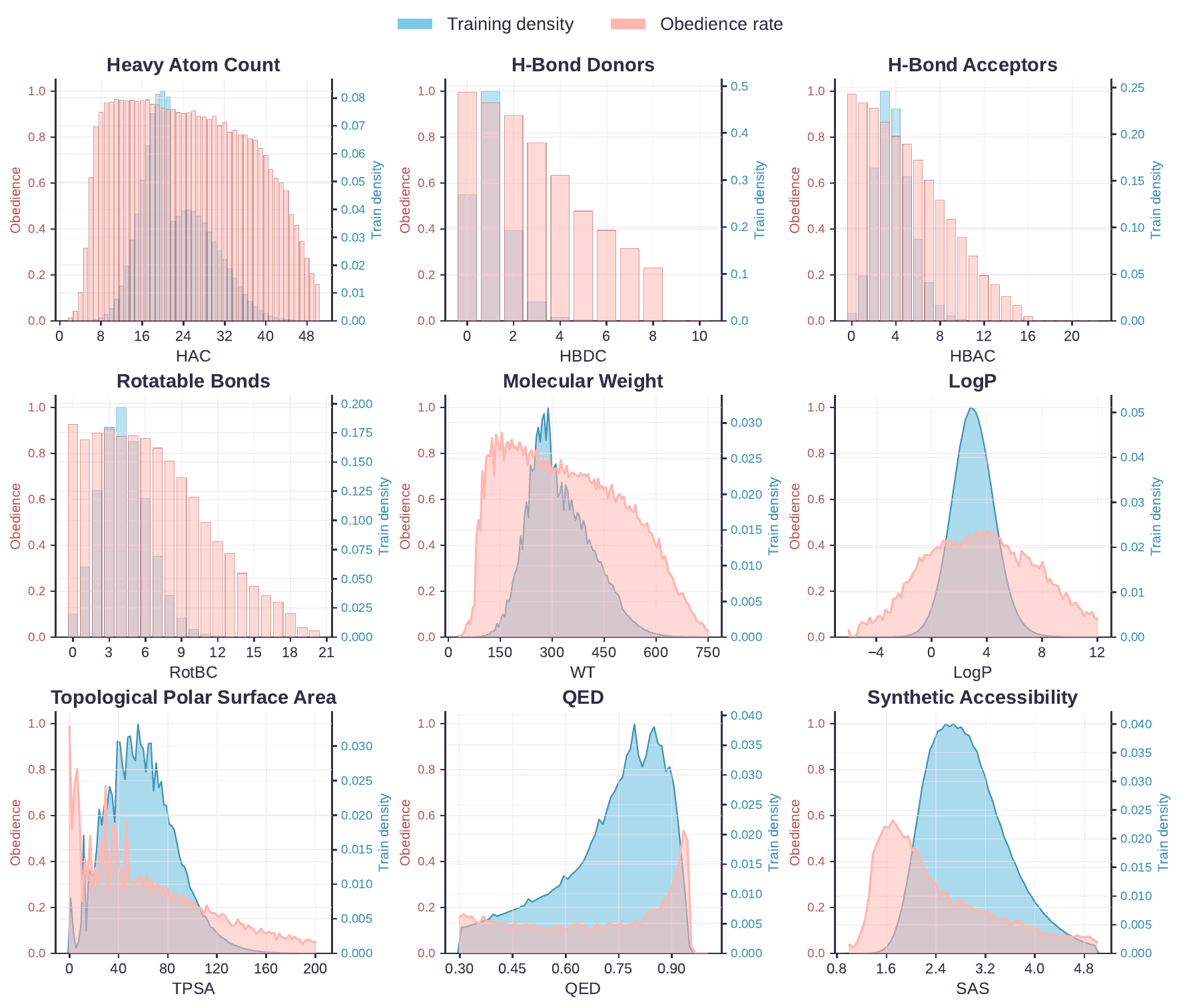}
  \end{center}
  \caption{
    \textbf{Single-property obedience rate versus training-set abundance for the nine
    conditioned properties.} In each panel, the pink bars show the obedience
    rate at each sampling point (left axis) and the green histogram shows the
    abundance of that property value in the training set (right axis).
  }
  \label{fig:obedience-abundance}
  \vskip -0.1in
\end{figure}

\Cref{fig:obedience-abundance} reports the obedience rate at each sampling point together with the abundance of the corresponding property value in the training set. The obedience rate is correlated with training-set abundance, but the conditioning also generalizes: relative to the densest region of the training data, \molexar{} retains a high obedience rate even at sampling points where training data are scarce.

\begin{table}[h]
  \caption{\textbf{Single-property conditioning configurations.} Discrete properties use
    one-hot encoding (exact match); continuous properties use radial basis
    function (RBF) encoding. The tolerance is the obedience window half-width:
    a generated molecule counts as matching when its property lies within
    $\pm$tolerance of the target.}
  \label{tab:property-config}
  \begin{center}
    \footnotesize
    \begin{tabular}{lllll}
        \toprule
        Property & Method & Range & Steps & Tolerance \\
        \midrule
        Heavy Atom Count (HAC)                & one-hot & $[2,\,50]$    & 49  & exact \\
        H-Bond Donors (HBDC)                  & one-hot & $[0,\,10]$    & 11  & exact \\
        H-Bond Acceptors (HBAC)               & one-hot & $[0,\,22]$    & 23  & exact \\
        Rotatable Bonds (RotBC)               & one-hot & $[0,\,20]$    & 21  & exact \\
        \midrule
        Molecular Weight (WT, Da)             & RBF     & $[30,\,750]$  & 128 & $\pm 5.67$  \\
        LogP                                  & RBF     & $[-6,\,12]$   & 96  & $\pm 0.19$  \\
        Topological Polar Surface Area (TPSA) & RBF     & $[0,\,200]$   & 96  & $\pm 2.11$  \\
        QED                                   & RBF     & $[0.3,\,1.0]$ & 64  & $\pm 0.011$ \\
        Synthetic Accessibility (SAS)         & RBF     & $[1.0,\,5.0]$ & 64  & $\pm 0.063$ \\
        \bottomrule
      \end{tabular}
  \end{center}
  \vskip -0.1in
\end{table}

\clearpage

\section{Correlation between Molecular Properties in the Training Set}
\label{sec:training-property-correlation}

\begin{figure}[h]
  \begin{center}
    \includegraphics[width=0.5\columnwidth]{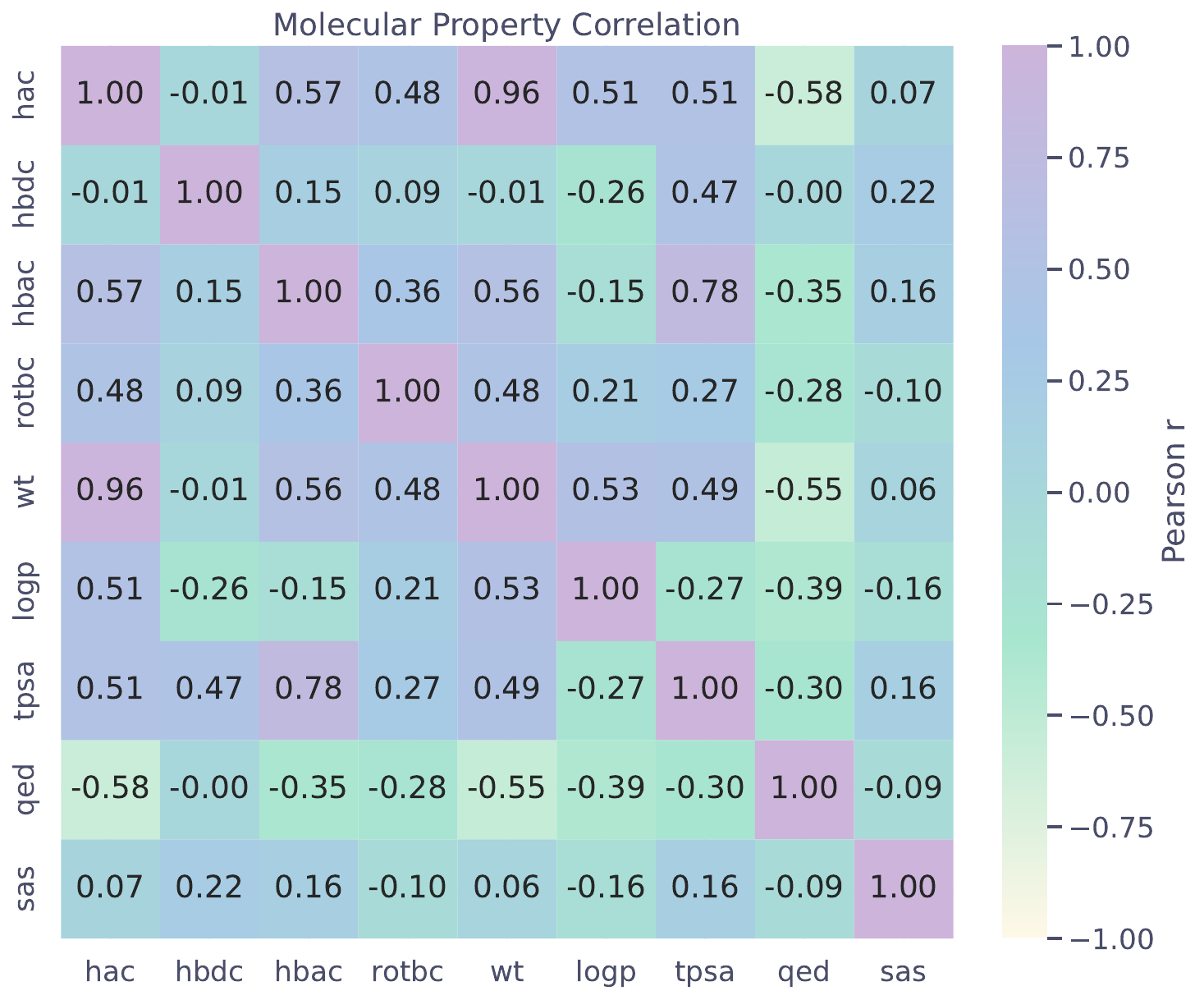}
  \end{center}
  \caption{
    \textbf{Pairwise Pearson correlation coefficients ($r$) between the nine
    conditioning properties in the training set.}
  }
  \label{fig:property-correlation}
  \vskip -0.1in
\end{figure}

Because different molecular properties can be correlated, and such correlations may affect the model's behavior under multi-property conditioning, we computed the pairwise Pearson correlation coefficients among the nine conditioning properties in the training set and report them in \cref{fig:property-correlation}. The figure shows the degree of correlation among the properties and provides context for interpreting the model's behavior under multi-property conditioning.

The correlations affect our chosen tasks as follows. HAC and RotBC are relatively strongly correlated ($r=0.48$), yet \molexar{} maintains largely independent control over the two in this setting, consistent with its strong obedience. QED and SAS are nearly uncorrelated ($r=-0.09$), indicating that they vary relatively independently in the training set. WT and LogP are more strongly correlated ($r=0.53$), and the positively correlated target values we chose in the corresponding test follow this trend.

\section{Target-Conditioned Generation on the CrossDocked2020 Benchmark}
\label{sec:target-crossdocked}

We provide the detailed per-metric results behind \cref{fig:target-radar} in \cref{tab:target,tab:pharmacophore}.

\begin{table*}[h]
  \begin{center}
  \begin{minipage}[t]{0.47\textwidth}
    \centering
    \captionof{table}{\textbf{Pocket- and sequence-conditioned generation results.}}
    \label{tab:target}
    \scriptsize
    \setlength{\tabcolsep}{2pt}
    \begin{tabular}{@{}l *{8}{c}@{}}
      \toprule
      \colhead \\
      \midrule
      \rowcolor{macaronblue}AR                        & 0.93 & \textbf{1.00} & 0.84 & 0.51 & 0.64 & 4.75 & -6.87 & 37.8 \\
      \rowcolor{macaronblue}Pocket2Mol                & 0.98$^\ast$ & \textbf{1.00} & 0.87$^\ast$ & 0.57 & 0.75 & 4.88$^\ast$ & -7.39 & 48.4 \\
      \rowcolor{macaronblue}DeepICL                   & \underline{0.99} & \underline{0.99} & 0.86 & \underline{0.61} & 0.33 & \textbf{4.94} & -7.32 & 48.3 \\
      \rowcolor{macaronblue}TargetDiff                & 0.92 & \underline{0.99} & \underline{0.89} & 0.49 & 0.60 & 4.58 & \underline{-7.70} & \underline{53.5} \\
      \rowcolor{macaronblue}DiffSBDD                  & \textbf{1.00} & \textbf{1.00} & \textbf{0.90} & 0.47 & 0.58 & 4.53 & -7.27 & 46.8 \\
      \rowcolor{macaronblue}PMDM                      & \underline{0.99} & \textbf{1.00} & \textbf{0.90} & 0.56 & 0.62 & 4.78 & -7.49$^\ast$ & 49.5 \\
      \rowcolor{macaronblue}DrugFlow                  & 0.85 & 0.96 & 0.84 & 0.51 & 0.71 & 4.52 & -7.18 & 46.3 \\
      \rowcolor{macaronblue}ResGen                    & 0.97 & \textbf{1.00} & 0.85 & 0.59$^\ast$ & 0.79$^\ast$ & \underline{4.91} & -6.58 & 34.6 \\
      \rowcolor{macaronblue}TamGen                    & \textbf{1.00} & \textbf{1.00} & 0.87$^\ast$ & 0.56 & 0.77 & 4.83 & -6.85 & 35.8 \\
      \rowcolor{macaronblue}PocketXMol                & 0.94 & 0.97 & 0.84 & 0.51 & 0.77 & 4.57 & \textbf{-7.81} & \textbf{57.6} \\
      \rowcolor{macaronblue}OMTRA (pocket)            & 0.94 & 0.97 & 0.87$^\ast$ & 0.50 & 0.77 & 4.32 & -6.86 & 40.6 \\
      \rowcolor{macaronpurple}PhoreGen (complex)        & 0.70 & 0.93 & 0.80 & 0.42 & 0.71 & 4.07 & -7.24 & 52.3 \\
      \rowcolor{macaronpurple}OMTRA (complex)           & 0.66 & 0.96 & 0.80 & 0.34 & 0.63 & 3.61 & -7.17 & 50.0 \\
      \midrule
      \molexar{}~(sequence)     & \textbf{1.00} & 0.98$^\ast$ & 0.83 & \textbf{0.65} & \underline{0.82} & 4.74 & -7.25 & 43.1 \\
      \molexar{}~(pocket)       & \textbf{1.00} & 0.97 & 0.84 & \textbf{0.65} & \textbf{0.83} & 4.82 & -7.42 & 53.0$^\ast$ \\
      \midrule
      Reference                 & 1.00 & - & - & 0.48 & 0.73 & 4.27 & -7.45 & - \\
      \bottomrule
    \end{tabular}
  \end{minipage}
  \hfill
  \begin{minipage}[t]{0.47\textwidth}
    \centering
    \captionof{table}{\textbf{Pharmacophore-conditioned generation results.}}
    \label{tab:pharmacophore}
    \scriptsize
    \setlength{\tabcolsep}{2pt}
    \begin{tabular}{@{}l *{8}{c}@{}}
      \toprule
      \colhead \\
      \midrule
      \rowcolor{macaronred}PGMG                      & \underline{0.94} & \textbf{1.00} & \underline{0.80} & \underline{0.50} & \underline{0.77} & 4.40$^\ast$ & \textbf{-7.23} & \underline{45.2} \\
      \rowcolor{macaronred}ShEPhERD                  & 0.81$^\ast$ & 0.78 & \underline{0.80} & 0.43$^\ast$ & 0.55 & \textbf{4.84} & -6.06 & 13.6 \\
      \rowcolor{macaronred}PhoreGen (pharmacophore)  & 0.64 & \underline{0.96} & \textbf{0.82} & 0.38 & 0.69 & 4.07 & \underline{-6.80} & \textbf{47.7} \\
      \rowcolor{macaronred}OMTRA (pharmacophore)     & 0.36 & 0.90 & 0.57 & 0.17 & \textbf{0.86} & 2.84 & -5.19 & 8.6 \\
      \midrule
      \molexar{}~(pharmacophore)& \textbf{1.00} & 0.91$^\ast$ & 0.76$^\ast$ & \textbf{0.59} & 0.71$^\ast$ & \underline{4.69} & -6.79$^\ast$ & 38.4$^\ast$ \\
      \midrule
      Reference                 & \textbf{1.00} & - & - & 0.48 & 0.73 & 4.27 & -7.45 & - \\
      \bottomrule
    \end{tabular}

    \vspace{3pt}
    {\scriptsize\raggedright \colorbox{macaronblue}{Blue}: pocket-only; \colorbox{macaronred}{Red}: pharmacophore-only; \colorbox{macaronpurple}{Purple}: complex (pocket+pharmacophore).\par}
    {\scriptsize\raggedright Valid.: Validity; Uniq.: Uniqueness.; Div.: Diversity; SA: Normalized synthetic accessibility; Lip.: Lipinski's rule of five; Vina: Vina docking score (docking values greater than 0 are excluded); HA: High affinity ratio.\par}
    {\scriptsize\raggedright Top 3 results are highlighted with \textbf{bold} text, \underline{underlined} text, and $^\ast$, respectively.\par}
  \end{minipage}
  \end{center}
  \vskip -0.1in
\end{table*}

\clearpage

\section{General Properties of \molexar{} on MolGenBench}
\label{sec:molgenbench-appendix}

\begin{figure}[h]
  \begin{center}
    \includegraphics[width=0.9\columnwidth]{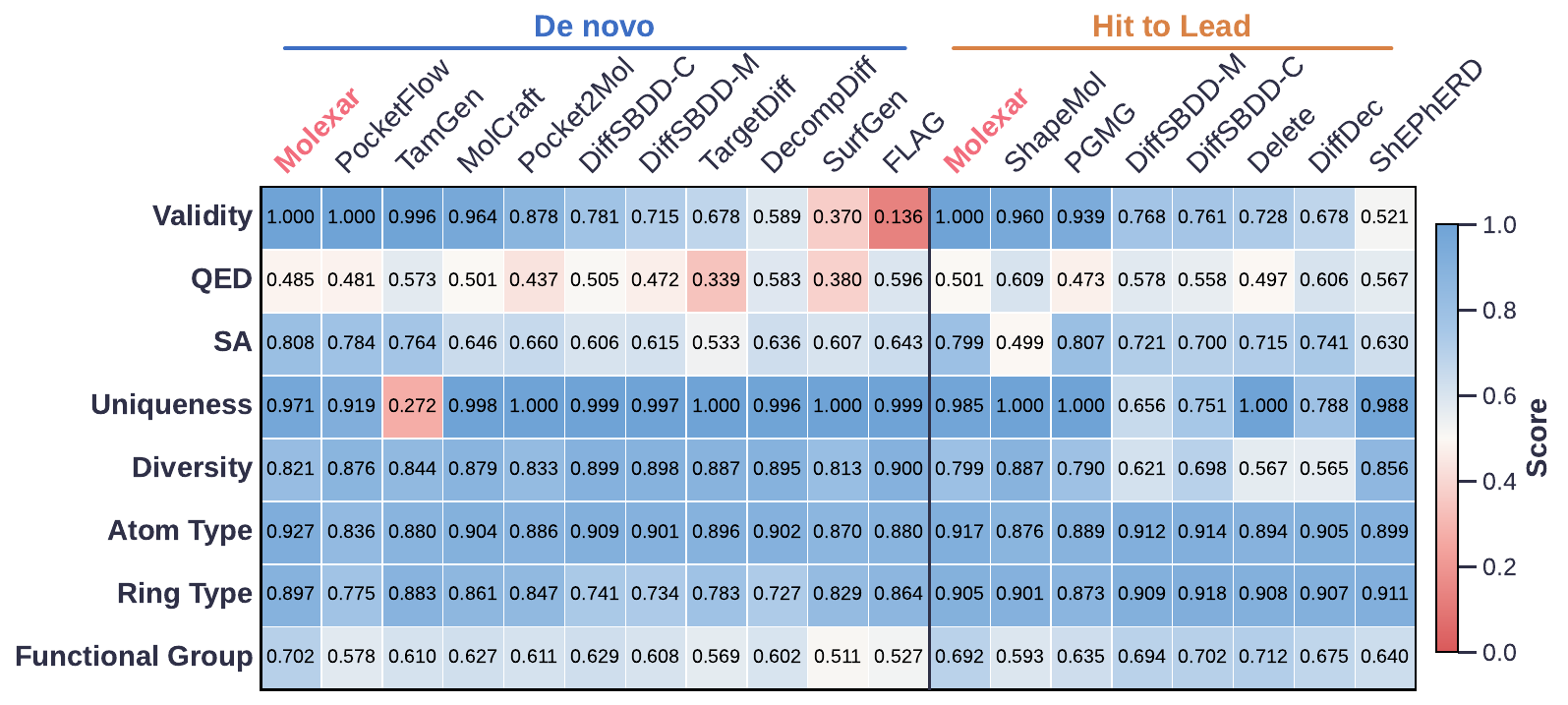}
  \end{center}
  \caption{
    \textbf{General properties of \molexar{} on MolGenBench.}
    Evaluation results of basic molecular properties, including validity, QED, SA score, uniqueness, diversity, and motif distributions
    (atom types, ring types, functional groups) with reference actives. Values represent the mean from three independent replicates; 
    all metrics are normalized to a 0--1 scale, where higher values indicate better performance.
  }
  \label{fig:molgenbench-properties}
  \vskip -0.1in
\end{figure}

\section{Examples of Generated Molecules}
\label{sec:molecule-showcase}

\begin{figure}[h]
  \begin{center}
    \includegraphics[width=0.8\columnwidth]{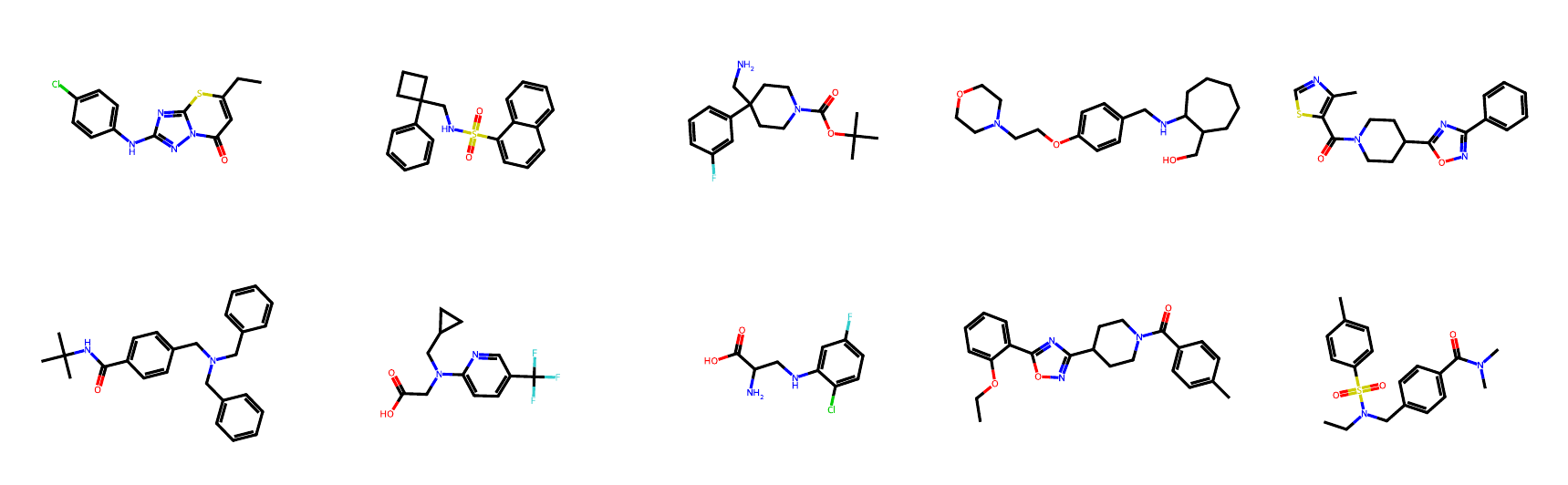}
  \end{center}
  \caption{
    \textbf{Examples of generated molecules on \textit{de novo} generation.}
  }
  \label{fig:unconditional-mols}
  \vskip -0.1in
\end{figure}

\begin{figure*}[h]
\centering
\footnotesize
\setlength{\tabcolsep}{6pt}
\renewcommand{\arraystretch}{1.0}
\begin{tabular}{m{4.4cm}|m{2.1cm}|m{8cm}}
\toprule
\centering Task & \centering Input & \centering\arraybackslash Generated molecules \\
\midrule
Linker design \& Scaffold morphing & \centering\includegraphics[height=1.1cm]{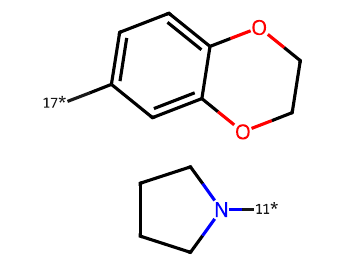}   & \centering\arraybackslash\includegraphics[height=1.4cm]{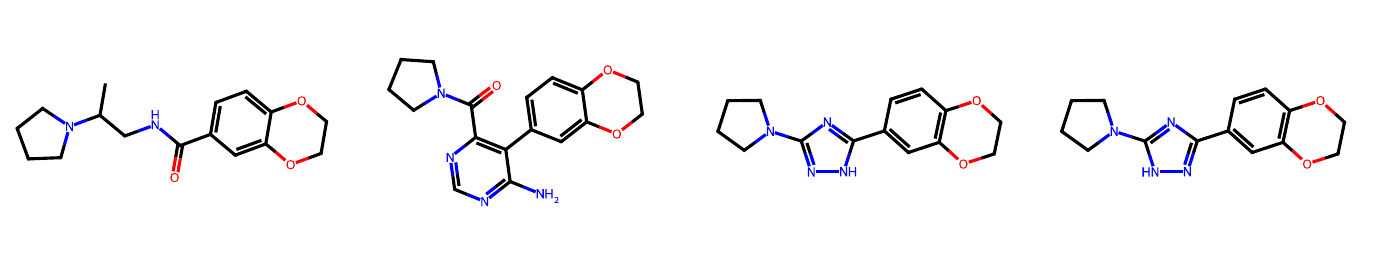}   \\
Motif extension                     & \centering\includegraphics[height=1.0cm]{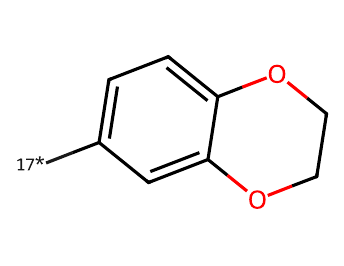}     & \centering\arraybackslash\includegraphics[height=1.4cm]{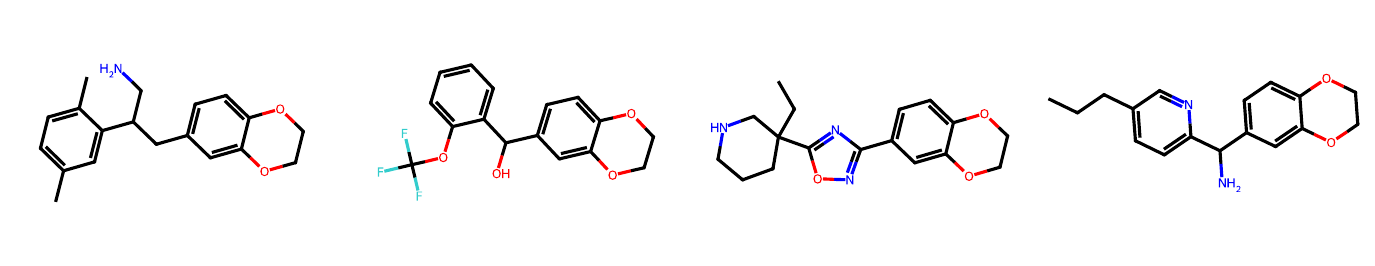}     \\
Scaffold decoration                 & \centering\includegraphics[height=1.4cm]{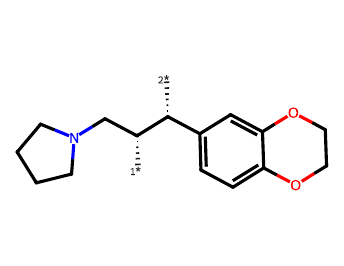} & \centering\arraybackslash\includegraphics[height=1.4cm]{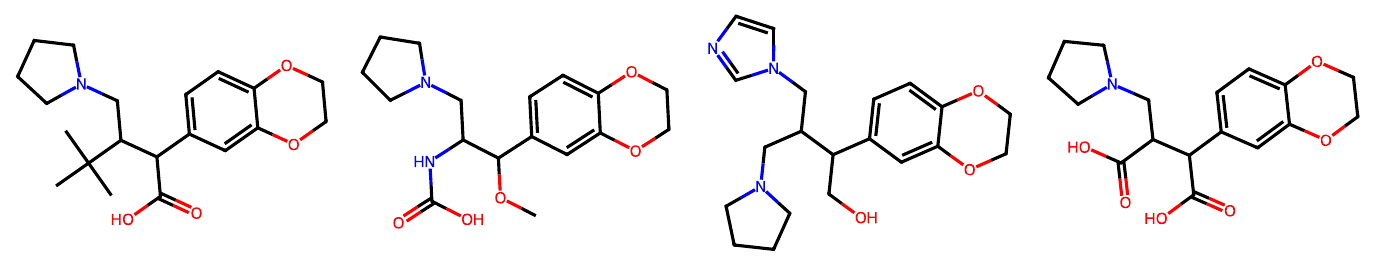} \\
Superstructure generation           & \centering\includegraphics[height=1.4cm]{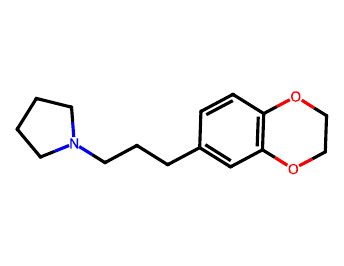}      & \centering\arraybackslash\includegraphics[height=1.4cm]{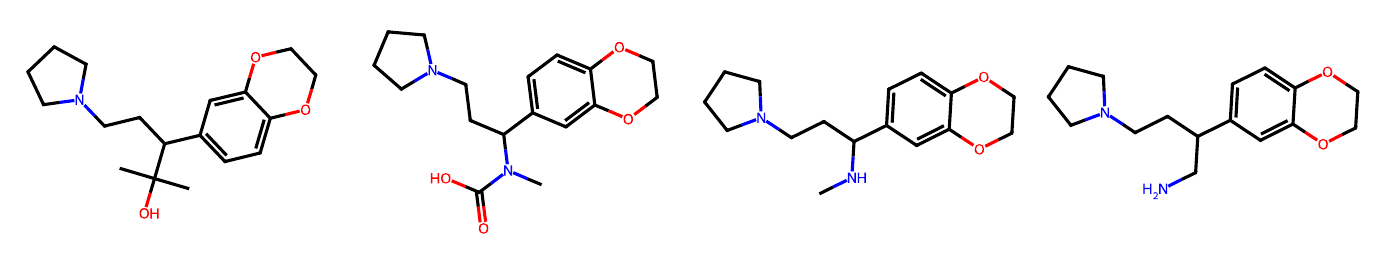}      \\
\bottomrule
\end{tabular}
\caption{\textbf{Examples of generated molecules on fragment-constrained generation of Eliglustat.}}
\label{fig:frag-constrained}
\vskip -0.1in
\end{figure*}

\begin{figure}[h]
  \begin{center}
    \includegraphics[width=0.7\columnwidth]{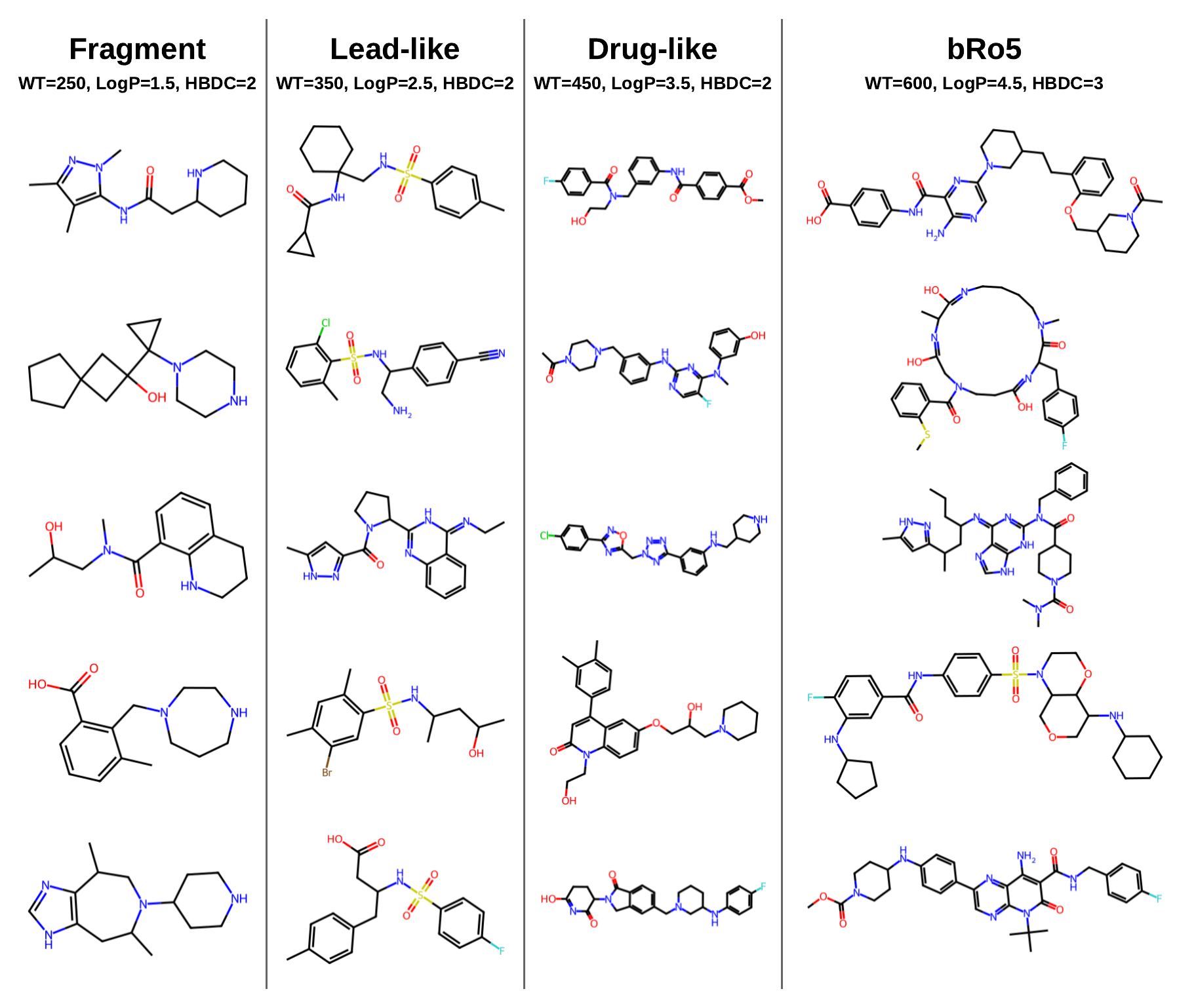}
  \end{center}
  \caption{
    \textbf{Examples of molecules generated under the realistic drug-regime
    multi-property settings.}
  }
  \label{fig:drug-mol}
  \vskip -0.1in
\end{figure}


\end{document}